\documentstyle[epsfig,12pt,a4wide]{article}

\def\bee{\begin{equation}}
\def\eee{\end{equation}}
\def\barr{\begin{array}}
\def\earr{\end{array}}
\def\ben{\begin{equation}}
\def\een{\end{equation}}
\def\bena{\begin{eqnarray*}}
\def\eena{\end{eqnarray*}}
\def\spa#1{\phantom{\fbox{\rule[-#1cm]{0cm}{0cm}}}}
\font\mybb=msbm10 at 12pt
\def\bb#1{\hbox{\mybb#1}}

\def\b1{e^0}

\def\lesssim{\mathrel{\hbox{\rlap{\hbox{\lower4pt\hbox{$\sim$}}}\hbox{$<$}}}}
\def\gtrsim{\mathrel{\hbox{\rlap{\hbox{\lower4pt\hbox{$\sim$}}}\hbox{$>$}}}}

\def\Ai{{\rm{Ai}}}

\begin{document}

\vspace*{-.6in} \thispagestyle{empty}
\begin{flushright}
CFP--2005--04\\
\end{flushright}
\vspace{.2in} {\Large
\begin{center}
{\bf Hagedorn transition and chronology protection \\in string theory}
\end{center}}
\vspace{.2in}
\begin{center}
Miguel~S.~Costa, Carlos~A.~R.~Herdeiro, J.~Penedones and N.~Sousa\footnote{Emails: miguelc, crherdei, jpenedones, nsousa@fc.up.pt}
\\
\vspace{.2in} 
\emph{ Departamento de F\'\i sica e Centro de F\'\i sica do Porto\\
Faculdade de Ci\^encias da Universidade do Porto\\
Rua do Campo Alegre, 687\\
4169-007 Porto, Portugal}
\end{center}

\vspace{.3in}

\begin{abstract}
We conjecture  chronology is protected in string theory due to the condensation of light winding strings near closed
null curves. This condensation triggers a Hagedorn phase transition, whose end-point target space geometry should be
chronological. Contrary to conventional arguments, chronology is protected by an infrared effect. We support this 
conjecture by studying strings in the O-plane orbifold, where we show that some winding string states are unstable and
condense in the non-causal region of spacetime. The one-loop string partition function has infrared divergences associated
to the condensation of these states.
\end{abstract}

\newpage

\section{Introduction}

The possibility of time travel has been one of the theoretical physicists' favorite puzzles. 
Of particular interest is time traveling to the past, since it raises many
causality paradoxes. Although our experience tells us such voyages are unlikely to be achievable, 
there is no rigorous proof of their impossibility. In an attempt to address
this problem, Hawking put forward a chronology protection conjecture \cite{Hawking:1991nk},
extending earlier results by Tipler  \cite{Tipler:1977eb}.
His argument was based on the behavior of quantum field theory in the presence of 
closed null curves, since in a spacetime with a Cauchy surface any future causality problems must be preceded by the 
formation of one such curve. Hawking argued that the 
one-loop energy-momentum tensor becomes very large near a closed null curve, 
therefore causing a significant backreaction in the geometry. The outcome would be a naked singularity 
or a gravitational evolution that prevents formation of the closed null curve itself. Either way, the mechanism 
that protects chronology is an ultraviolet effect. Since the ultraviolet behavior of string theory is different from 
that of quantum field theories, it is questionable that the above mechanism will protect chronology.

Theories with extended objects provide privileged probes of spacetimes with closed causal curves since the extended objects can wrap around them.
As mentioned above, closed null curves are particularly important because causality problems first arise when such curves are created.
In the framework of string theory, there are winding string states that can become light just before they wrap a closed null curve,
when their proper length is of the string scale. A phase transition can then occur, analogous to the Hagedorn phase transition 
where winding string states become massless for a compactification circle with radius of the string scale \cite{Hagedorn:1965st,Atick:1988si}. 
It is therefore 
plausible that light winding states play a role in protecting chronology. In this work we establish this analogy in detail by studying a toy model,
where we are able to show that such phase transition does in fact occur and it is associated to the condensation of string fields 
in the causally problematic region.
We expect this mechanism to hold in general, which leads us to the following conjecture: {\em closed null curves do not form in string theory because
light winding states condense, causing a phase transition whose end-point target space geometry is chronological}. This
mechanism for protecting chronology is truly stringy and is an infrared effect, in sharp contrast with Hawking's proposal.

The toy model we shall consider is an orbifold of three-dimensional Minkowski space introduced in \cite{reviewcosta}, 
called the O-plane orbifold. The orbifold quotient space is a good laboratory to study the issue of chronology
protection because it is free of singularities and it develops closed causal curves. There is, however,
a truncation of spacetime that is free of closed causal curves. It is the main point of this work to
show that some winding string states, which condense in the pathological region of spacetime, give a possible
mechanism for the protection of chronology. For other works of strings on orbifolds with closed causal curves and on 
closed string backgrounds with bad chronology see [6--47].

Since the orbifold closed causal curves are topological, it is questionable 
whether the conjectured mechanism is generic, in the sense that it might not work for non-topological closed causal curves dictated
by the Einstein equations \cite{Carter:1968rr}. However, when fermions obey anti-periodic boundary conditions around
the closed curves, we believe the mechanism is general because  the appearance of unstable
winding states relies only on the existence of closed curves with very small proper length.

We start our discussion in section two by giving a detailed review of the O-plane orbifold. In particular, we analyze its causal structure, 
wave functions and quantum field theory divergences. We find infrared divergences associated to the condensation of some particle 
states and show that the theory is ultraviolet divergent because of the presence of the non-causal region. In section three
we consider strings in the O-plane orbifold. Again we analyze the string wave functions and compute the one-loop
partition function, which when appropriately interpreted has only infrared divergences. These divergences are associated to the presence of 
unstable string states, whose condensation is conjectured to protect chronology in section four. 
We give some concluding remarks in section five. Appendix A contains some technical results 
regarding wave functions and corresponding Hilbert space measure. In appendix B we give an alternative derivation
of the particle and string partition functions using the path integral formalism.

\section{Review of O-plane orbifold} \label{review}

\begin{figure}
\centering\epsfig{file=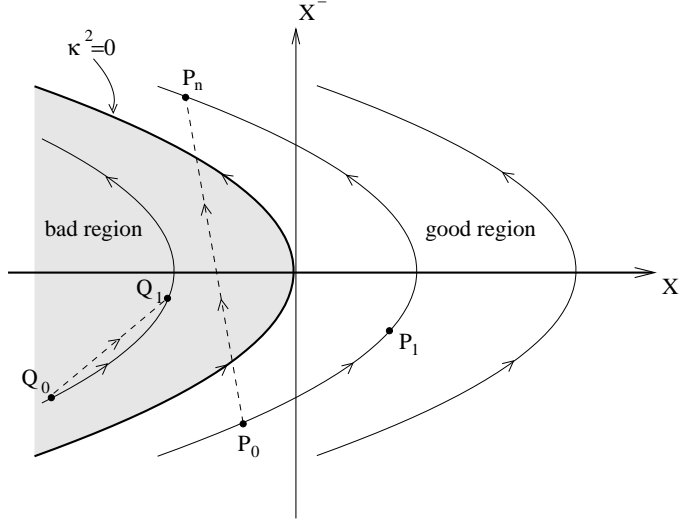,width=9cm}
\caption{\small{Orbits of the Killing vector field $\kappa$ projected in the $XX^-$--plane of the covering space. 
The surface $\kappa^2=0$ divides spacetime in a good region and a bad region.
The dashed straight lines are timelike geodesics connecting image points, projected in the $XX^-$--plane.}}
\label{figure1}
\end{figure}

In this section we shall review some basic facts about the O-plane orbifold introduced in \cite{reviewcosta}. 
This is an orbifold of three-dimensional Minkowski spacetime. To define it 
consider the light-cone coordinates $(x^-,x,x^+)$, normalized such that the flat  metric is given by
$ds^2=-2dx^+dx^-+dx^2$. The orbifold group element $\Omega = e^{\kappa} $ 
is generated by the Killing vector
$$
\kappa=2\pi i\left(R P_- + \Delta J \right) \ , 
$$
where 
$$ 
    iJ   = x_+ \frac{\partial\ }{\partial x} - x \frac{\partial\ }{\partial x^+}\ , \ \ \ \ \ \ 
    iP_- =     \frac{\partial\ }{\partial x^-} \ , 
$$
are, respectively, the generators of a null boost and a null translation and $R$ and $\Delta$ are 
constants.
The orbits of this Killing vector field can be found in figure \ref{figure1}.
The orbifold identification can be explicitly written as
$$ 
    \vec{x}\,\equiv\,\left( \barr{l} x^-
    \\ x \\ x^+ \earr \right) \ \ \sim \
    \ \left( \barr{c} x^- + 2\pi R \spa{0.1}\\ \displaystyle{x -2\pi\Delta x^- -2\pi^2 R\Delta} \spa{0.1}\\
    \displaystyle{x^+-2\pi\Delta x +2\pi^2\Delta^2 x^- + \frac{4}{3}\,\pi^3 R\Delta^2} \earr \right) \ . 
$$ 

Since the generator of the O-plane orbifold involves the time direction
one might wonder about the causal structure of quotient space, and in particular about the 
existence of closed causal curves. For this purpose it is convenient to introduce a new set 
of coordinates $\vec{y}=(y^-,y,y^+)$, in which the orbifold action becomes trivial. 
These are defined by 
$$ 
\barr{l} x^-=\,y^- \\
\displaystyle{x^{\phantom{-}}=\,y-\frac{E}{2}( y^-)^2} \\
\displaystyle{x^+=\,y^+-Eyy^-+\frac{E^2}{6}(y^-)^3}\ , 
\earr 
$$ 
where $E\equiv \Delta/R$. Now the orbifold identification becomes
simply 
$$ 
y^-\sim y^-+2\pi R
$$ 
and the metric 
$$
ds^2=-2dy^-dy^++2Ey(dy^-)^2+dy^2\ .
$$
The surface $y=0$ divides spacetime in two regions, one with $\kappa$ spacelike ($y>0$) -- the good region -- and the
other with $\kappa$ timelike ($y<0$) -- the bad region. The region with $y<0$ is expected to be
pathological. Indeed, although there are closed timelike curves passing through every point in this space, these curves always
enter the $y<0$ region as shown in figure \ref{figure1}.  
Moreover, in the covering space there are points which are light-like separated from their image.
These points lie on the so-called {\em polarization surfaces} \cite{Kim:1991mc}. 
Null curves joining a point and its image become closed null curves in quotient space and 
are dangerous in quantum
field theory because they are responsible for extra divergences, as we shall review in subsection 2.3. 

\subsection{Particle dynamics}

To understand dynamics in the O-plane orbifold, we start by analyzing the motion of point particles. 
Clearly, particle trajectories are straight lines in covering space. However, we are 
interested in studying the problem from the view point of an observer using $\vec{y}$-coordinates, 
which reduces to considering the following Lagrangian
$$
{\mathcal{L}}=-2\dot{y}^+\dot{y}^-+2Ey(\dot{y}^-)^2+\dot{y}^2 \ .
$$
Dots are derivatives with respect to an affine parameter. We now introduce the canonical conjugate momenta $(p_-,p,p_+)$, of which the first 
and third components are conserved. This is because the quotient space preserves only two of the six isometries of three-dimensional 
Minkowski space, namely those generated by $\partial_+ \equiv \partial/\partial{y^+}$ and $\partial_- \equiv \partial/\partial{y^-}$. 
The problem then simplifies to the motion of the particle along direction $y$, with an effective Hamiltonian given by
\bee
    {\cal H}=p^2+V(y) \ , \ \ \ \ \ \ \ \ \ V(y)\equiv -2Ep_+^{\,2}y-2p_+p_- \ .  
\label{HV}
\eee
The motion is thus determined by a linear potential. The constant force associated to this potential is clearly an inertial force, 
since the $\vec{y}$-coordinates 
describe an accelerating frame. 

Now comes an important point. The Hamiltonian, which is a constant of motion, is
the momentum squared of the particle
\bee
    {\cal H} = p_\mu p^\mu = - M^2 \ , 
\label{defmass}
\eee
and is therefore fixed by the particle's mass. 
The classical turning point $y_0$ for a particle of light-cone energy $p_+$ and Ka\l u\.{z}a-Klein momentum $p_-$ is then found
by setting $p=0$ in the previous equation
\bee
    M^2 = 2 p_+ p_- + 2Ey_0p_+^{\,2}   \ . 
\label{particlemassshell}
\eee

\begin{figure}[t]
\centering\epsfig{file=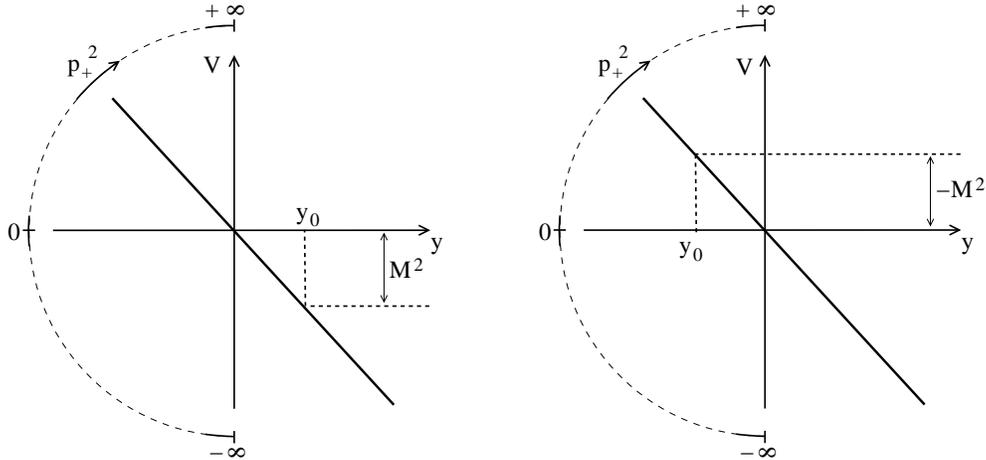,width=13cm}
\caption{\small{Effective potential $V(y)$ and conserved energy $-M^2$ for particle motion along direction $y$. Shown are the cases
of $M^2>0$ (left) and $M^2<0$ (right) for $p_-=0$. The turning point $y_0$ is defined by $V(y_0)=-M^2$. 
Particle states with $p_+^{\,2}<0$ are allowed quantum mechanically.}}
\label{figure2}
\end{figure}

Consider first a particle of $M^2 > 0$ and $p_-=0$. It is clear from figure \ref{figure2} that such a particle will never 
enter the  bad region $y<0$.
Since the slope of the potential gets steeper as $p_+$ increases, the higher the light-cone energy of the particle, 
the closer it gets to $y=0$. In the low energy limit $p_+\rightarrow 0$, 
the particle is pushed away to $y=+\infty$. On the other hand, particles of $M^2 < 0$ and $p_-=0$ always cross into the bad region and, for 
low energies, reach very far out into this region as can be seen from figure  \ref{figure2}.

For particles with $p_- \neq 0$ the potential is displaced along $y$, so it no longer goes through 
$y=0$. However, the above qualitative behavior is the same: particles with $M^2>0$ are repelled away from the bad region and 
for low energies are pushed away to $y=+\infty$; particles with $M^2<0$ are also repelled but for low energies 
reach far out into the bad region.

\subsection{Single particle wave functions}

\begin{figure}[t]
\centering\epsfig{file=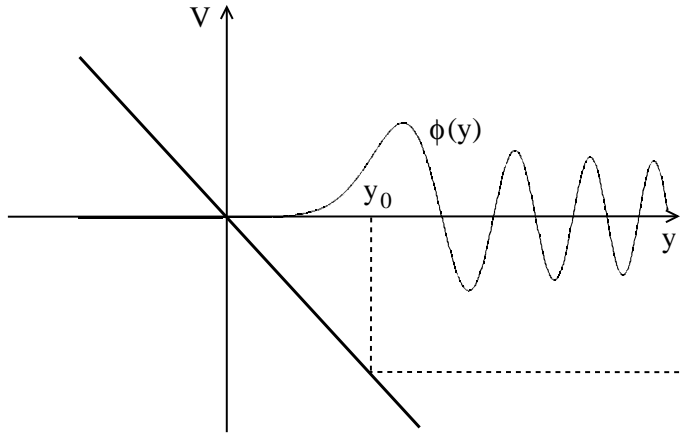,width=9cm}
\caption{\small{Profile of a wave function in the $y$-direction.}}
\label{figure3}
\end{figure}

To quantize the particle, we follow the usual prescription of replacing $p_\mu \rightarrow -i D_\mu$ 
in the mass-shell condition (\ref{defmass}). The single particle wave functions $\phi(y^-,y,y^+)$ obey the Klein-Gordon equation
$$
    \Big( -2 {\partial_+}\partial_- -2Ey \, {\partial_+^{\, 2}} + \partial_y^{\, 2} \, \Big)\, \phi = M^2\,\phi \ .
$$
To find the solutions to this differential equation we first consider a complete basis of functions in the orbifold. 
The details are presented in the appendix A.1; here we state the necessary results. In addition to the orbifold periodicity $0\le y^-\le 2\pi R$, 
we consider a large box defined by $0\le y^+ \le L_+$ and  $-L/2\le y \le L/2$. The wave functions
\bee
\phi_{p_+,y_0,m}(\vec{y}) = \frac{|K(p_+)|^{1/3}}{\sqrt{2\pi RL_+ \rho(y_0,p_+)}}\,\Ai(z)\,e^{\,i\left(p_+y^+ + \frac{m}{R}y^-\right)} \ ,
\label{wavefunctions}
\eee
form a complete normalized basis and are eigenfunctions of the Laplacian in the O-plane orbifold. The Ka\l u\.{z}a-Klein momentum is
quantized as $p_- = m/R$ and the other quantum numbers are the light-cone energy $p_+$ and the classical turning point $y_0$.
The function ${\rm Ai}(z)$ is an Airy function with argument given by $z^3 =K (y_0-y)^3$ and
\bee
K =K (p_+)= 2Ep_+^{\,2}\ ,\ \ \ \ \ \ 
\rho(y_0,p_+) = \frac{1}{\pi}\,\left[ \,K(p_+) \left({\rm sgn}(K)\,\frac{L}{2} - y_0\right) \, \right]^{1/2}\ .
\label{K}
\eee
A $y$-plot of a wave function is given in figure \ref{figure3}.
To sum over states in the Hilbert space the appropriate integration measure is given by
$$
\sum_{m=-\infty}^{\infty}\, \frac{L_+}{2\pi}\,\int_{-\infty}^{\infty} dp_+ \,\int_{-\frac{L}{2}}^{\frac{L}{2}} dy_0\,\rho(y_0,p_+)\ .
$$

On-shell wave functions are eigenfunctions of the Laplacian with eigenvalue $M^2$ and therefore
their quantum numbers satisfy the mass-shell condition (\ref{particlemassshell}). To define the 
evolution of a given field configuration, we shall consider an initial value problem defined
on a surface of constant $y^+$. The normal $\tilde{n}$ to this surface has norm given by
$\tilde{n}^2 = -2Ey$.
Thus, in the good region $y>0$ this surface is space-like and the initial value
problem is standard. However, for $y<0$ we are specifying the ``initial'' data
on a time-like surface. We shall interpret this ``initial'' data as a boundary
condition. This is satisfactory to understand the evolution of perturbations
in the good region.  In appendix A.2 we construct a basis of functions
on a surface of constant $y^+$. The light-cone evolution of
each mode will be  determined by its light-cone energy fixed by the on-shell relation.
In the appendix we show that, in order to be normalizable on a surface of constant $y^+$,
the modes must have $K\in \bb{R}$. In particular, for $K<0$, we have unstable modes with $p_+$ purely 
imaginary and $y_0$ complex.

To understand the subtleties of matter propagation in the O-plane orbifold, we consider the cases of massive particles ($M^2 > 0$) 
and tachyons $(M^2 < 0)$ separately, both for $m=0$. 
Massless particles can be included in the case of massive particles by introducing an infrared regulator. 

\begin{figure}[t]
\centering\epsfig{file=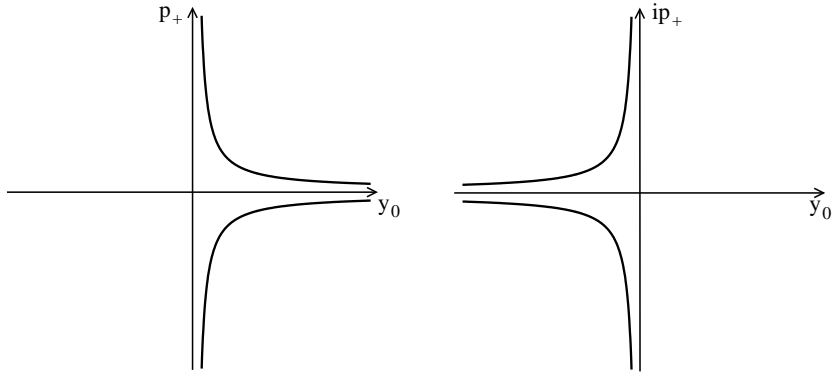,width=11cm}
\caption{\small{Dispersion relation for particles with $M^2>0$ and $m=0$.}}
\label{figure4}
\end{figure}

For massive particles, the behavior of the wave function can be easily understood from the corresponding classical effective potential in 
(\ref{HV}). The turning point $y_0$ is now the point where the wave function turns from oscillatory to evanescent (see figure \ref{figure2}). 
For on-shell particles, the mass formula (\ref{particlemassshell}), with $p_-=0$, defines the light-cone energy for a given $y_0$. 
This dispersion relation is plotted in figure \ref{figure4}. 
As expected from the classical behavior, in the low energy limit we have $y_0 \rightarrow +\infty $ and the 
wave function vanishes everywhere.
In appendix A.2 we show that the set of on-shell wave functions with  $y_0\in \bb{R}$ is a complete basis of
normalizable functions on a surface of $y^+$ constant.  Thus, 
from figure \ref{figure4}, we see that to include modes with $y_0<0$ we must allow for imaginary $p_+$. 
Then we are effectively dealing with the classical potential $V(y)$ with $p_+^{\,2}<0$ and the wave functions oscillate in the 
bad region. Since the light-cone energy is imaginary, these modes can grow in time.
A generic field configuration will then grow in light-cone time $y^+$, mainly in the bad region.  
We conclude that massive particles have unstable modes in this spacetime.
If the bad region were to be excised, we would expect to be left only with well-behaved modes.

\begin{figure}[t]
\centering\epsfig{file=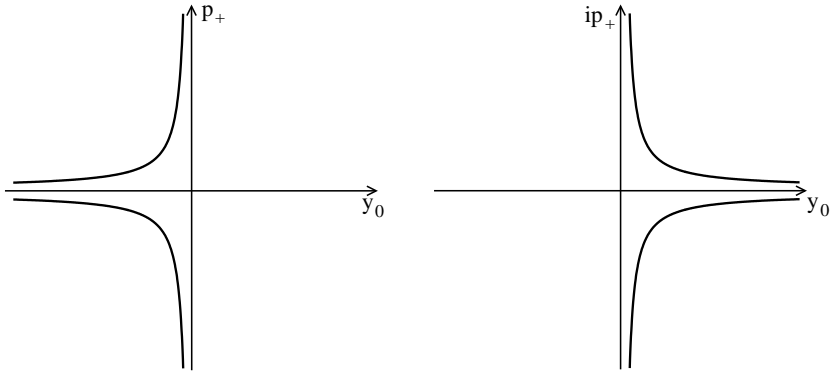,width=11cm}
\caption{\small{Dispersion relation for particles with $M^2<0$ and $m=0$.}}
\label{figure5}
\end{figure}

Next consider a tachyon, also with $m=0$. As shown in figure \ref{figure5}, for $p_+$ real the turning point 
lies at $y_0<0$ and the wave function oscillates according to the classical trajectory in figure \ref{figure2}. 
To allow for states with $y_0>0$ one must consider imaginary $p_+$. These unstable modes oscillate in space for
$-\infty<y<y_0$ and therefore reach far out into the good region. 
Contrary to massive particles, in the low energy limit the tachyon wave function is a plane wave throughout space, 
which is the usual result for tachyons in Minkowski space.

Particles with non-vanishing Ka\l u\.{z}a-Klein momentum have a similar qualitative behavior. In particular, the condition
for having on-shell stable states with real $y_0$
can be obtained from the mass formula (\ref{particlemassshell})
\bee
2Ey_0\,\left( M^2 + \frac{m^2}{R^2(y_0)}\right) > 0\ ,
\label{unstpart}
\eee
where $R^2(y_0) = 2Ey_0R^2$ is the proper radius squared of the orbifold compact direction, computed at the turning point $y_0$.
When this condition is not satisfied, $p_+$ becomes complex and the associated wave functions are not normalizable
on a surface of constant $y^+$. 
However, in appendix A.2, we show that there are other unstable on--shell states, with $p_+$ purely imaginary
and $y_0$ complex, which are normalizable.
 
\subsection{Quantum field theory and ultraviolet catastrophe}

To compare with the string theory results presented later on in this paper, we study the usual quantum field theory pathologies
in the O-plane orbifold due to the presence of closed causal curves. These are the ultraviolet problems that led Hawking to conjecture that a strong 
gravitational backreaction would protect chronology. 

Consider first the existence of polarization surfaces, arising from light-like separated images. Since
the covering space is Minkowski space, it is simple to compute the relativistic interval between image points 
$$
\Delta_n^{2}\equiv 
\|\vec{x}-\vec{x}_n\|^2 = 8E(\pi nR)^2\left(y-\frac{E}{6}\,(\pi nR)^2\right)\ .
$$
In particular, points located at
$$
y=y_{n}\equiv\frac{E}{6}\,(\pi nR)^2 \ , 
$$
are null separated from their $n$-th image. It is important to note that the corresponding closed null curve 
intersects the $y=0$ surface, so  it is possible that cutting off the $y<0$ region cures any pathology 
arising from the polarization surfaces, as proposed in \cite{Cornalba:2002nv,Cornalba:2003ze}.

For a scalar field $\phi$ of mass $M$ the two-point function can be written in Euclidean Schwinger parameterization as
$$
\langle \phi(\vec{x}) \phi(\vec{x}\,') \rangle = \sum_w \frac{e^{-M \Delta_w}}{4\pi\Delta_w}
=\sum_w \int_0^{+\infty}\frac{dl}{2}\,\frac{1}{(2\pi l)^{3/2}}\,\exp\left(-\frac{l}{2}\,M^2 - \frac{\Delta^2_w}{2l}\right) \ ,
$$
where $ \Delta_w^2 = \|\vec{x}-\vec{x}_w\,\!\!\!'\|^2$ and the trajectory winding number $w$ runs from $-\infty$ to $+\infty$.
When $\vec{x}$ lies on the $n$-th polarization surface, the two-point 
function has extra divergences as $\vec{x}\,' \rightarrow \vec{x}$, coming from the $w=\pm n$ terms in the sum. These are ultraviolet
divergences that are not renormalizable \cite{Kim:1991mc}. Likewise, the expectation value of the energy-momentum tensor
diverges at  the polarization surfaces. Again, these problems could be cured if one excises the bad region. 
It is the main point of this paper to suggest that a truly stringy phenomena in fact excises such region, hence protecting chronology.

Next consider the one-loop contribution to the vacuum energy, written as 
$$
Z=\int_0^{+\infty} \frac{dl}{2l}\;{\rm {\bf Tr}} \;\exp\left[-i\,\frac{l}{2}\left(M^2 + p_\mu p^\mu \right)\right]\ ,
$$
where the trace is over off-shell single particle states.
This is written in first quantized language so as to compare it later with the string torus amplitude. 
Choosing the basis $|p_+,y_0,m\rangle$ to perform the trace and Wick rotating $l\rightarrow -il$, 
$p_+\rightarrow ip_+$, we arrive at the expression
$$
Z= i \int_0^{+\infty} \frac{dl}{2l}\,\sum_m\, \frac{L_+}{2\pi}\,
\int dp_+ \, dy_0\,\rho(y_0,ip_+)\;
\exp\left[-\frac{l}{2}\left( M^2 - 2ip_+\frac{m}{R} + 2Ey_0p_+^{\,2}\right)\right]\ .
$$
One could question the validity of this Wick rotation since the final expression clearly has large $l$ 
divergences that were absent in the original one. The alternative would be to choose a state dependent 
Wick rotation depending on the sign of $(M^2+p_\mu p^\mu)$. However, in string theory the requirement 
of modular invariance forces us to treat all modes equally \cite{Craps:2002ii}. The chosen Wick rotation
is physically reasonable because massive particle states that propagate in the good region, i.e. states with $y_0>0$,
give a finite contribution to the partition function. On the other hand, states with $y_0<0$ will originate
infrared divergences provided $p_+$ is large enough, for both massive and tachyonic particles.

Restricting to states with $y_0>0$ and for large $l$, the $p_+$ integral can be done using the  
saddle point approximation. The $l$ integral is then dominated by an exponential with the following
argument
$$
-\frac{l}{2}\left( M^2  + \frac{m^2}{R^2(y_0)}\right)\ .
$$
Hence, the condition for this integral to be infrared convergent is equivalent to the condition for the existence of
on-shell stable states with $y_0$ positive, as can be seen from (\ref{unstpart}). For massive particles, this condition
holds for all on-shell states that oscillate in the good region, which have $y_0>0$. On the other hand, for tachyonic 
particles, the above condition is not always satisfied and therefore the partition function becomes infrared divergent. 
This is related to the existence of on-shell unstable states. These states oscillate in both good and bad regions
and therefore their condensation would change the whole of spacetime.

To analyze the ultraviolet behavior of the partition function it is useful to consider the expression obtained from the path integral 
formalism. We report on this computation in appendix B and on its derivation starting from the above canonical result in appendix A.1. 
Here we simply give the final result
$$
Z=iL_+(2\pi R)\sum_w\,\int dy\, \int_0^{\infty} \frac{dl}{2l}\,\frac{1}{(2\pi l)^{3/2}}\;
\exp\left(-\frac{l}{2}\, M^2 - \frac{\big(2\pi w R(y)\big)^2}{2l}\right)\ ,
$$
where $y$ is now the particle's average position along this direction and $w$ is the winding number of closed trajectories.
The $w=0$ term gives the usual renormalizable ultraviolet divergence from the region $l\rightarrow 0$. The other terms  are ultraviolet 
finite for $y>0$, diverge for $y\rightarrow 0$ and acquire, by analytic continuation, an imaginary part for $y<0$. 
These are extra ultraviolet divergences that are absent in string theory.

\section{Strings in the O-plane orbifold} \label{classicalstring}

We now move on from particles to strings. We work with the bosonic string in critical dimension,  adding
the transverse space $\bb{R}^{23}$ to the O-plane orbifold. Since these directions are mere spectators
we include their contribution explicitly only when necessary.
We work in units where $\alpha'=1$. 

When considering strings the novelty is the appearance of twisted sectors, which correspond to strings winding around the 
$y^-$-direction. As we shall see, these winding states play a crucial role in protecting chronology. 

Start with the string Lagrangian
\bee
{\mathcal{L}}=\frac{1}{2\pi}\left[-\dot{Y}^+\dot{Y}^-+EY(\dot{Y}^-)^2
+\frac{\dot{Y}^2}{2}+(Y^+)'(Y^-)'-EY(Y^-)'^2-\frac{Y'^2}{2}\right] \ , 
\label{lagrangestring}
\eee
where dots and primes denote derivatives of the worldsheet fields $(Y^-,Y,Y^+)$ with respect to the worldsheet coordinates 
$\tau$ and $\sigma$, respectively. The  coordinate $\sigma$ runs from 0 to $\pi$.
Consider the center of mass motion of a closed string that winds $w$ times around the compact direction. The embedding functions have the form 
\bee
    Y^-(\tau,\sigma)= 2R w \sigma +y^-(\tau) \ , \ \ \ \ \ \ Y(\tau,\sigma)=y(\tau) \ , \ \ \ \ \ \ \ Y^+(\tau,\sigma)=y^+(\tau) \ , 
\label{yclass}
\eee
from which we derive the string center of mass conjugate momenta
\bee
p_- = -\frac{1}{2} \, (\dot{y}^+ - 2E y \dot{y}^-) \ , \ \ \ \ \ \ p = \frac{1}{2} \, \dot{y} \ , \ \ \ \ \ \ p_+ =- \frac{1}{2}\, \dot{y}^- \ .
\label{conjugatemomenta}
\eee
Again, only $p_+$ and $p_-$ are conserved. As for the particle case, the motion along $y$ is determined by the effective Hamiltonian 
\bee
    {\cal H} =  p^2 + V(y) \ , \ \ \ \ \ \ 
    V(y) \equiv - 2E\left( p_+^{\,2}-\left(w R\right)^2\right)y- 2p_+p_- \ . 
\label{stringhamiltonian}
\eee
We see that there is an additional linear contribution to the potential from the winding of the string. Its interpretation is simple. Since
the proper length of the compact direction increases with ${y}$, it costs energy for a winding string to increase its value of $y$. 
Defining the conserved quantity $\lambda = - {\cal H}$, the string center of mass turning point $y_0$, where $p=0$,  satisfies
\bee
    \lambda = 2p_+p_- + 2Ey_0\left( p_+^{\, 2} - \left(w R\right)^2 \right) \ . 
\label{stringdispersion}
\eee
For the particular solution (\ref{yclass}) the Hamiltonian  in (\ref{stringhamiltonian})
generates translations along the worldsheet time and it is
related to the Virasoro zero-modes by ${\cal H} = 2( L_0 + \tilde{L}_0)$. This is zero by the classical Virasoro constraints. 
As we shall see in subsection 3.2, in the quantum theory
$2( L_0 + \tilde{L}_0) = {\cal H} + 2 (N+\tilde{N}) = 4$, where $N$ and $\tilde{N}$ are the usual number operators.
For each string state it is therefore necessary to consider both negative and positive values 
of $\lambda = 2(n+\tilde{n}-2)$. 
The corresponding center of mass classical dynamics reduces to the motion of a particle of constant energy in the potential in 
(\ref{stringhamiltonian}). We analyze it below, together with the associated wave functions.

\subsection{Winding states dynamics and wave functions}

We have just seen that on-shell classical string states satisfy the relation  $\lambda = - {\cal H}$, which using (\ref{stringhamiltonian})
yields
\bee
   - p_\mu p^\mu =  \lambda + 2E y \left(w R\right)^2 \ . 
\label{lambdaoperator}
\eee
This constraint defines the wave equation for the string center of mass
$$
    \Big( -2 {\partial_+}\partial_- -2Ey \, {\partial_+^{\, 2}} + \partial_y^{\, 2} \, \Big)\, \phi = M^2(y)\, \phi \ ,
$$
where 
\bee
M^2(y) =  \lambda + 2E y \left(w R\right)^2 \ . 
\label{stringmass}
\eee
This Klein-Gordon equation can be derived from the Lagrangian
$$
{\cal L} = \partial \phi \partial \bar{\phi} + M^2(y) \phi\bar{\phi} 
$$
for a complex scalar field associated with strings of winding number $\pm w$. 
The major difference to the particle case is that this string mass depends on the $y$-coordinate and therefore it is not conserved. 
From a pure field theoretical view point one
would then expect that something special happens for $M(y)=0$. We shall come back to this point in section 4.
 
Solutions to the above differential equation have the same form as for the particle case. The only difference is that $K$ in the expression 
(\ref{wavefunctions}) for the basis of functions and in the corresponding definitions (\ref{K}) is now given by
$$
K(p_+) = 2E\left(p_+^{\,2} - \left(w R\right)^2\right)\ .
$$ 
On-shell states satisfy the string dispersion relation (\ref{stringdispersion}). As for the particle case, on-shell
wave functions, which are normalizable on  a surface of constant $y^+$,  have $K\in \bb{R}$. 
For $K<0$ this corresponds to unstable modes with $p_+$ purely imaginary and $y_0$ complex.

\begin{figure}[t!]
\centering\psfig{file=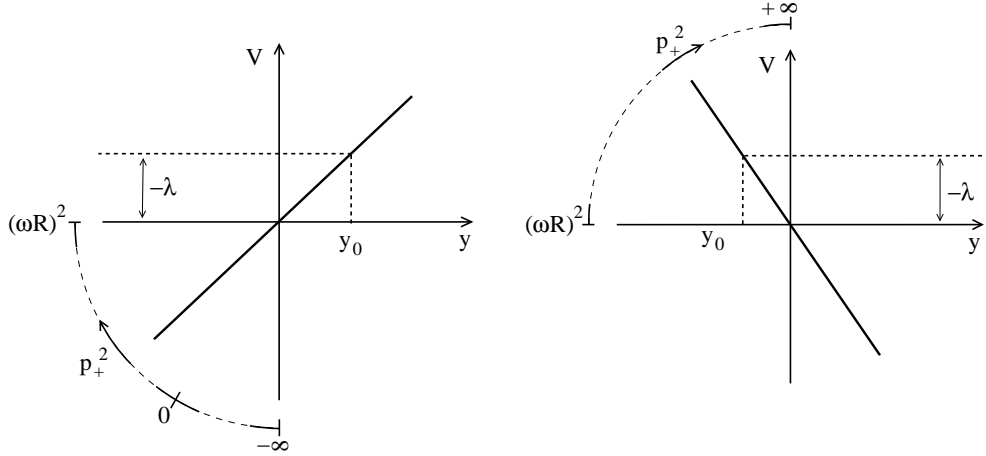,width=13cm}
\caption{\small{Effective potential $V(y)$ and conserved energy $-\lambda$ for string motion along direction $y$. Shown is the case
 of $\lambda<0$ and $m=0$, for $p_+^{\,2}<(w R)^2$ (left) and $p_+^{\,2}>(w R)^2$ (right). 
For $p_+^{\,2}=(w R)^2$ the potential becomes flat.
In contrast with the particle case, 
when $p_+=0$ the  potential is tilted.}}
\label{figure6}
\end{figure}

\begin{figure}[b!]
\centering\epsfig{file=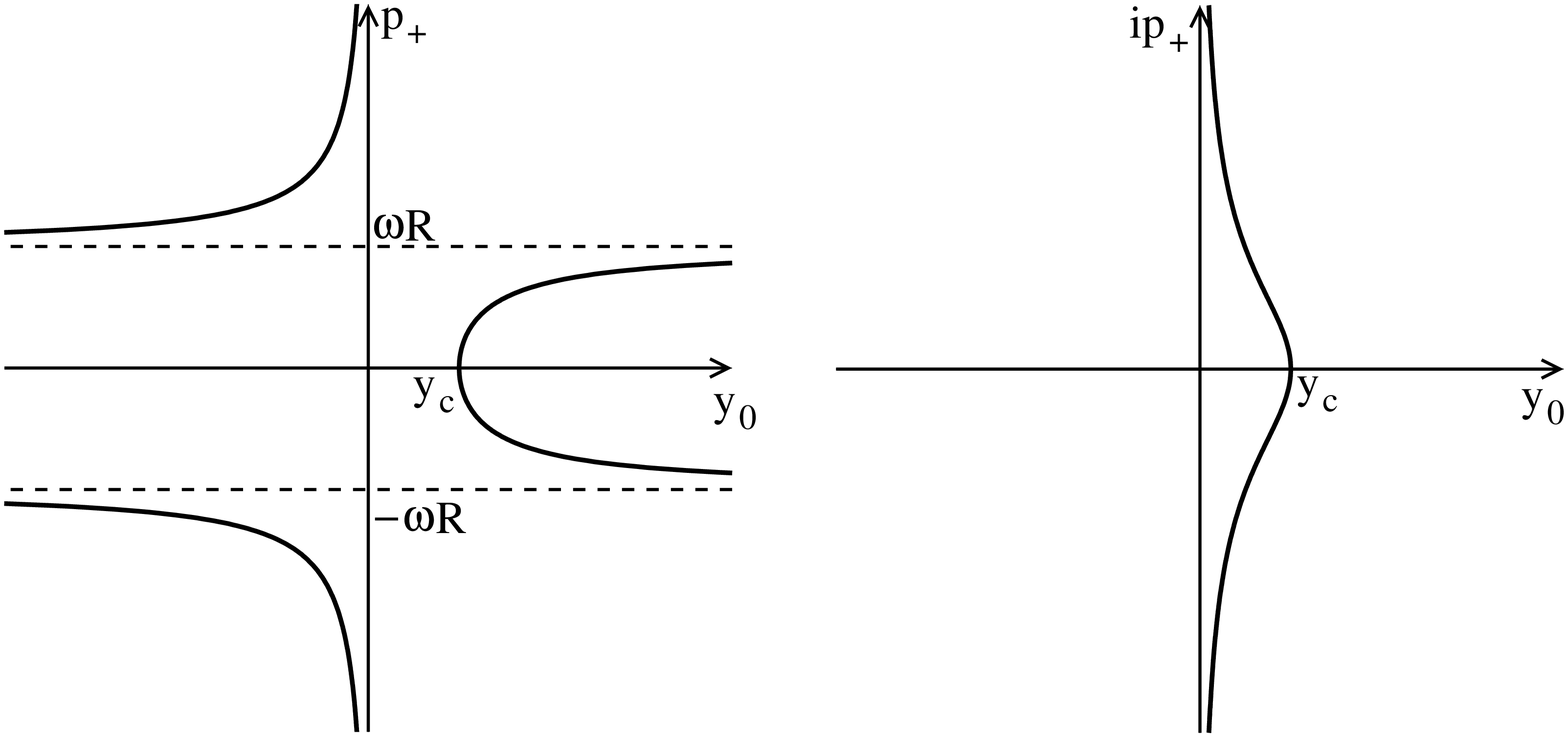,width=11cm}
\caption{\small{Dispersion relation for string states with $\lambda<0$ and $m=0$.}}
\label{figure7}
\end{figure}

Let us first focus on the case $\lambda<0$ and, for simplicity, $m=0$. 
The classical trajectories and behavior of wave functions  can be immediately understood from figure \ref{figure6}. There are three different 
kinematic regimes, determined by the light-cone energy $p_+$. For $0 \leq p_+^{\,2} < (w R)^2$ the string comes from $y=-\infty$
until it reaches a turning point $y_0>0$, where it bounces back. The wave function oscillates up to the turning point and becomes 
evanescent thereafter. In the low energy limit $p_+=0$, the slope of potential remains positive and the wave function oscillates
for $y<y_0$. This oscillatory behavior is analogous to a tachyon in Minkowski space, whose wave function also has spatial oscillations 
for low energies. We shall come back to this important issue later because these states will be crucial in protecting chronology.
As one increases $p_+$, the slope decreases and vanishes for
$p_+^{\,2} = (w R)^2$. At this point the string moves at constant velocity, reflecting a 
cancellation between the string tension and the inertial force due the accelerating $\vec{y}$-frame. The corresponding wave function is a plane wave. 
Finally, when $ p_+^{\,2} > (w R)^2$ the string comes from $y=+\infty$ and turns around at $y_0<0$. 
Its wave function becomes evanescent beyond that point.
For large light-cone energies, the potential becomes infinitely steep, as it was the case for particles.

\begin{figure}[t!]
\centering\epsfig{file=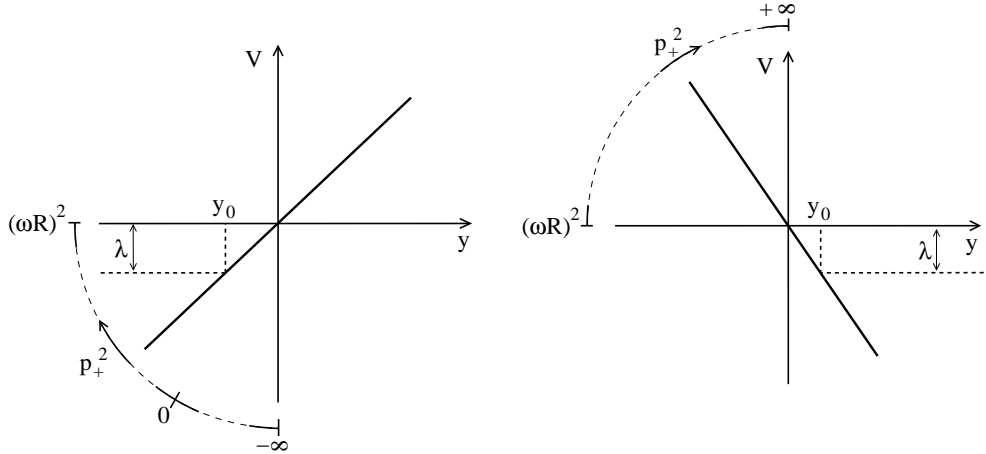,width=13cm}
\caption{\small{Effective potential $V(y)$ and conserved energy $-\lambda$ for string motion along direction $y$. Shown is the case
of $\lambda>0$ and $m=0$, for $p_+^{\,2}<(w R)^2$ (left) and $p_+^{\,2}>(w R)^2$ (right). 
For $p_+^{\,2}=(w R)^2$ there is no classical trajectory nor wave function.
In contrast with the particle case, 
when $p_+=0$ the effective potential is tilted.}}
\label{figure8}
\end{figure}

In figure \ref{figure7} we plot the dispersion relation for states with $\lambda<0$ and $m=0$. 
As shown in appendix A.2, to determine the evolution of generic perturbations defined on a surface of
constant $y^+$, one must include all real values of $y_0$ in the
on-shell basis. Hence, we must consider states with 
imaginary $p_+$, which grow in time.  The turning point $y_0$ for these unstable states satisfies 
$0<y_0 < y_c(w)$ with
$$
y_c(w) = \frac{-\lambda}{2E(w R)^2}\ .
$$
One then expects that the condensation of these modes will drastically change spacetime in the region
$y<y_0$, where the wave function oscillates (see figure 6, left).

\begin{figure}[t!]
\centering\epsfig{file=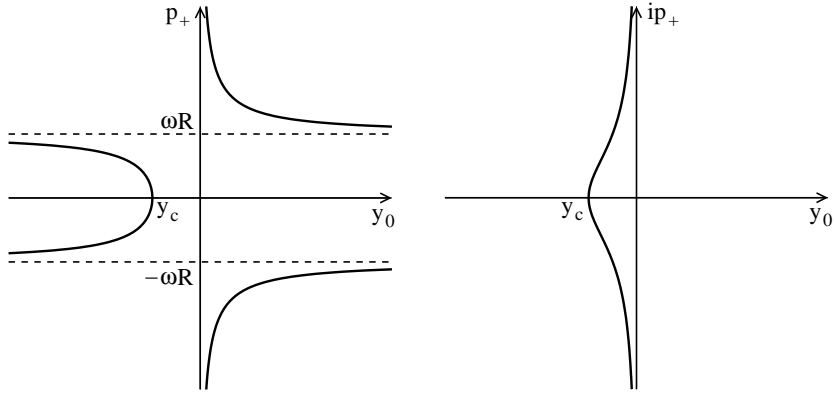,width=11cm}
\caption{\small{Dispersion relation for string states with $\lambda>0$ and $m=0$.}}
\label{figure9}
\end{figure}

We move to the $\lambda>0$ case, and again set $m=0$. Trajectories and wave functions can be understood looking at figure \ref{figure8}. 
For $0 \leq p_+^{\,2} < (w R)^2$ the string comes from negative $y$ and turns back at $y_0<0$. 
The low energy  limit $p_+=0$ is similar to the above case but with $y_0<0$. As the potential
tilts and one reaches $p_+^{\,2} = (w R)^2$, the turning point $y_0$ is pushed to $-\infty$ and the wave function disappears. For 
$p_+^{\,2} > (w R)^2$ the string comes in from positive $y$ and bounces back at $y_0>0$. The high energy limit is the same as before. 
In figure \ref{figure9} we plot the dispersion relation for $\lambda>0$ and $m=0$, both for $p_+$ real and imaginary. The unstable 
modes will grow considerably for $y<y_0<0$.

For a fixed value of $\lambda$ and generic quantum number $m$ not all real values of $y_0$ give rise to normalizable wave functions
on a surface of constant $y^+$. The condition for having on-shell stable states with real $y_0$ can be determined from the quadratic 
equation (\ref{stringdispersion})
\bee
2Ey_0\,\left( \lambda + \frac{m^2}{R^2(y_0)} + w^2R^2(y_0)\right) > 0\ ,
\label{condcond}
\eee
where $R^2(y_0)= 2Ey_0R^2$ is again the proper radius squared of the orbifold compact direction, 
computed at the turning point $y_0$. On-shell unstable states, which are normalizable on a surface of constant $y^+$,
occur for $p_+$ purely imaginary and $y_0$ complex fixed by the dispersion relation (\ref{stringdispersion}).

\subsection{Canonical quantization}\label{canonical}

We now quantize the string using $\vec{y}$-coordinates and in light-cone gauge. The worldsheet theory with the Lagrangian 
(\ref{lagrangestring}) is not a free field theory. However, the worldsheet field $Y^-$ obeys the free wave equation. Thus
the light-cone gauge can be implemented by demanding that $Y^-$ does not have oscillatory terms.
In this gauge, the solutions of the wave equation for the worldsheet fields 
have the following expansion
$$
    \barr{rcl}
        \displaystyle{Y^-(\tau,\sigma)}&=& \displaystyle{2Rw\sigma + y^-(\tau)}
           \spa{0.3}\\
        \displaystyle{Y(\tau,\sigma)^{\phantom{-}}}&=&
        \displaystyle{y(\tau) +\frac{i}{\sqrt{2}}\,\sum_{n \neq 0}\,\frac{1}{n}\,
                      \left(\tilde{a}_n\, e^{\,2 i n(\tau + \sigma)}+ a_n\,e^{\,2 i n(\tau - \sigma)}\right)}
           \spa{0.5}\\
        \displaystyle{Y^+(\tau,\sigma)}&=&
        \displaystyle{y^+(\tau) +\frac{i}{\sqrt{2}}\,\sum_{n \neq 0}\,\frac{1}{n}\,
                      \left(\tilde{a}_n^+(\tau)\, e^{\,2 i n(\tau+\sigma)}+ a_n^+(\tau)\,e^{\,2 i n(\tau - \sigma)}\right)} \ .
    \earr 
$$
As usual in light-cone gauge, the coefficients $\tilde{a}_n^+(\tau)$  and $a_n^+(\tau)$ are not independent degrees of freedom
because they are fixed by the Virasoro constraints $\tilde{L}_n=L_n=0$ for $n\ne 0$.
From the string Lagrangian (\ref{lagrangestring}) we obtain the conjugate momenta
$$
    P_- = -\frac{1}{2\pi}\,\left(\dot{Y}^+-2E Y \dot{Y}^-\right) \ , \ \ \ \ \ \ P = \frac{1}{2\pi}\,\dot{Y} \ , \ \ \ \ \ \ 
    P_+ = -\frac{1}{2\pi}\,\dot{Y}^- = - \frac{1}{2\pi}\, \dot{y}^- \ ,
$$
which can be integrated along $\sigma$ to yield the string center of mass momentum $p_\mu$ derived in (\ref{conjugatemomenta}).
The worldsheet energy-momentum tensor $T_{\pm\pm} = \partial_\pm Y^\mu \partial_\pm Y_\mu$ enables us to compute the Virasoro zero-modes
$$
    \barr{l}
        \displaystyle{L_0 = \frac{1}{4\pi} \int_0^\pi T_{--} d\sigma 
                          = \frac{1}{4}\, p_\mu p^\mu + \frac{1}{2}\,Ey \left(w R\right)^2 + \frac{mw}{2} + N} \ ,
           \spa{0.4}\\
        \displaystyle{\tilde{L}_0 = \frac{1}{4\pi} \int_0^\pi T_{++} d\sigma
                                  = \frac{1}{4}\, p_\mu p^\mu + \frac{1}{2}\,Ey \left(w R\right)^2 - \frac{mw}{2} + \tilde{N}} \ ,
    \earr 
$$
where $N$ and $\tilde{N}$ are the usual light-cone oscillator contributions. The Virasoro zero-modes sum and difference are
$$
    \barr{l}
        \displaystyle{L_0 + \tilde{L}_0 = \frac{1}{2}\,p_\mu p^\mu + Ey \left(w R\right)^2 + N + \tilde{N}} \ ,\spa{0.3}\\
        \displaystyle{L_0 - \tilde{L}_0 = {mw} + N - \tilde{N}} \ .
    \earr 
$$
Given any $N$ and $\tilde{N}$, studying the classical dynamics of the string center of mass reduces 
to the effective Hamiltonian problem described in subsection 3.2.
The corresponding conserved quantity $\lambda$ is fixed by setting $L_0=\tilde{L}_0=0$.

Canonical quantization of the string is done by imposing equal-time commutation relations. This gives the non-vanishing
commutators for the string center of mass
$$
[y^-(\tau), p_-(\tau)] = i \ ,\ \ \ \ \ \  [y(\tau), p(\tau)] = i \ ,\ \ \ \ \ \  [y^+(\tau), p_+(\tau)] = i 
$$
and for the oscillators 
$$
[a_n,a_m] = n\,\delta_{n+m}\ ,\ \ \ \ \ \ [\tilde{a}_n,\tilde{a}_m] = n\,\delta_{n+m}\ .
$$
Since the oscillators decouple from the zero modes the normal ordering for the Virasoro zero-modes is the usual one. For the bosonic
string one has $L_0=\tilde{L}_0=1$ and physical states satisfy the on-shell operator relation
$$
             M^2(y) \equiv - p_\mu p^\mu =  2Ey \left(w R\right)^2 + 2\left(N + \tilde{N} - 2\right) \ ,
$$
and the level matching condition
$$
             N - \tilde{N} = mw\ .
$$
The quantum numbers for the string center of mass wave functions $\phi_{p_+,y_0,m}$  will then obey the on-shell relation
$$
    \lambda  = 2 p_+\frac{m}{R} + 2Ey_0 \left(p_+^{\,2} -\left(w R\right)^2\right) = 2\left(n + \tilde{n}-2\right)\ . 
$$
We conclude that for the bosonic string both positive and negative values of $\lambda$ are allowed. The corresponding behavior
of the string center of mass wave functions was studied in the previous subsection, where we saw that states with negative
values of $\lambda$ grow significantly in the bad region, as well as a bit inside the good region. 

\subsection{Partition function}

We now compute the bosonic string partition function and find divergences associated with the condensation of physical
states. We shall see these are all infrared divergences.
In the canonical formalism one takes the trace over the Hilbert space of states
$$
    Z = \int_{\cal F} \frac{d^2 \tau}{\tau_2} \; {\rm {\bf Tr}}\left( q^{\,L_0-1}\bar{q}^{\,\tilde{L}_0-1} \right) \ ,
$$
where $q^{2\pi i \tau}$ and $\tau=\tau_1 + i\tau_2$ is the modular parameter of the Euclidean torus, 
restricted to the fundamental domain ${\cal F}$. Since we are computing the trace in  light-cone gauge, 
one integrates over all the string zero-modes but keeps only the physical
transverse oscillators. Using the explicit form of the Virasoro zero-modes it follows that
$$
    Z = \int_{\cal F} \frac{d^2 \tau}{\tau_2} \; {\rm {\bf Tr}}\, 
         \exp\left[ 2\pi i \tau_1 (N-\tilde{N}-mw) 
          +2\pi\tau_2 \left(-\frac{1}{2}\,p_\mu p^\mu - Ey\left(w R\right)^2 - N - \tilde{N} + 2\right)\right] \ .
$$
In the expression for the Virasoro zero-modes the string center of mass momentum and the number operators commute. 
Therefore one can perform the trace over the oscillators to obtain the same result as for Minkowski space.
Integrating over momenta in the spectator directions and choosing the basis $|p_+,y_0,m\rangle$ to perform the trace
over the remaining zero-modes we arrive at the expression
\bee
Z = L_+V_{23}\int_{\cal F} \frac{d^2 \tau}{\tau_2} \, {\cal Z}_0(\tau)\,
\sum_{m,w}\,\int \,\frac{dp_+}{2\pi} \, dy_0\,\rho(y_0,p_+)\;
\exp\Big(- 2\pi i \tau_1 mw +  \pi \tau_2  \lambda\,\Big)\ ,
\label{partitionfunction}
\eee
where ${\cal Z}_0 = (2\pi\sqrt{\tau_2})^{-23} |\eta(\tau)|^{-48}$ and we recall that on the basis $|p_+,y_0,m\rangle$ the operator 
${\cal H}$ defined in (\ref{stringhamiltonian}) has eigenvalue $-\lambda$ given by (\ref{stringdispersion}).

\subsubsection{Infrared divergences}

We want to understand the divergences in the partition function coming from the infrared region $\tau_2\rightarrow +\infty$. In particular, we 
are interested in investigating the contribution of each string state to such divergence. For this purpose we cut the ultraviolet region
of modular integration by setting $\tau_2 \ge 1$, in order to integrate over $\tau_1$ to enforce the level matching condition. Recalling 
the expansion of the Dedekind eta function $\eta(\tau)$ we have
$$
{\mathcal{Z}}_0=(2\pi \sqrt{\tau_2})^{-23}(q\bar{q})^{-1}\sum_{n,\tilde{n}=0}^{\infty}d_nd_{\tilde{n}}q^n\bar{q}^{\tilde{n}} \ , 
$$
where $d_n$ and $d_{\tilde{n}}$ represent the degeneracy of left and right moving states. As for the particle case we define the analytic 
continuation of the $p_+$ integral as $p_+\rightarrow ip_+$. The infrared behavior of the partition function
can then be studied from 
$$
\begin{array}{rcl}
    Z &\sim& \displaystyle{i\,\frac{L_+V_{23}}{(2\pi)^{23}} \sum_{m,w,n,\tilde{n}} d_nd_{\tilde{n}} 
    \,\int\,\frac{dp_+}{2\pi} \, dy_0\,\rho(y_0,p_+)\,\int_1^{+\infty} \frac{d \tau_2}{(\sqrt{\tau_2})^{25}}}\spa{0.5}\\
    && \displaystyle{\exp\left[ - 2 \pi \tau_2 \left( - ip_+ \frac{m}{R} + Ey_0\left( p_+^{\,2}+ \left(w R\right)^2 \right)
                                                      + n+\tilde{n} - 2 \right)\right]} \ ,
\end{array}
$$
where the sum is restricted by the level matching condition for physical states. The infrared divergences come from states
in the Hilbert space such that the real part of the argument in the above exponential 
is positive. Hence, given $(m,w,n,\tilde{n})$ and negative $y_0$,  the integral is infrared divergent
for $p_+$ large enough. Let us then focus on the region of integration with $y_0$ positive and look for further infrared divergences.  
Using the saddle point approximation we obtain  the large $\tau_2$ behavior of the 
$p_+$ integral. Then one is left with a  $\tau_2$ integral dominated by an exponential with argument
$$
-  \pi \tau_2  \left( 2\left(n+\tilde{n} -2\right)+\frac{m^2}{R^2(y_0)} + w^2 R^2(y_0)\right)\ .
$$
As for the particle case, we see that the condition for the integral to be infrared convergent is equivalent to the 
condition for the existence of on-shell stable states with $y_0$ positive, as can be seen from (\ref{condcond}).
For $\lambda>0$, this condition holds for all on-shell states that oscillate in the good region, which have $y_0>0$.
String states with $\lambda<0$, do not always satisfy the above condition and therefore give
rise to infrared divergences. In particular, the state with $m=0$, $w=1$ and $n=\tilde{n}=0$
renders the partition function divergent for $y_0<2/(ER^2)$. Excluding the usual closed string tachyon, 
this is the state that condenses the furthest into the $y>0$ region.

\subsubsection{Ultraviolet finiteness}

Up to now we have cut the ultraviolet region of modular integration to understand the
infrared origin of the divergences. In string theory one expects no further divergences arising from the ultraviolet. Let 
us analyze the partition function without cutting off the ultraviolet region. 
The divergences we will find correspond to those found in the previous analysis. 

Consider the expression  (\ref{partitionfunction}) for the partition function. 
Performing the Wick rotation, the sum in Ka\l u\.{z}a-Klein momentum can be Poisson re-summed using
$$
    \sum_m \exp{\left[ 2\pi i m \left(\tau_1 w -\frac{\tau_2 p_+ }{R}\right)\right]} = 
    \sum_{w'} \delta\left(\tau_1 w -\frac{\tau_2 p_+ }{R}-w' \right) \ ,
$$
such that the $p_+$ integral becomes trivial. As explained in appendix A.1, we can then use the integral representation of the measure
to interchange the integral over $y_0$ with the integral over the string center of mass $y$, arriving at the final expression 
$$
   Z = i\,\frac{L_+(2\pi R)V_{23}}{(2\pi)^{26}}\, \int_{\cal F} \frac{d^2\tau}{\tau_2^{14}}\,|\eta(\tau)|^{-48}\,\sum_{w,w'}\, 
       \int dy\; \exp\left(-\frac{\pi}{\tau_2}\,R^2(y)\,\bb{T}\bar{\bb{T}} \right)\ ,  
$$
where $\bb{T} = w \tau - w'$. In appendix B we derive independently this result using the  path integral formalism.
It is easy to check that the integrand is modular invariant. Unlike the particle case, for fixed $(w,w')$ and 
$y\rightarrow 0$ there are no ultraviolet divergences because the $\tau_2$ integration is restricted to the fundamental domain ${\cal F}$.

\subsubsection{Hagedorn behavior}

The term in the partition function with $w=w'=0$ gives the usual Minkowski space contribution. 
The remaining terms in the sum are the contribution from the winding states and Ka\l u\.{z}a-Klein states in their dual guise. 
From now on we consider only these terms since they are the ones which contain the orbifold information. 
Using the usual trick of replacing the fundamental region ${\cal F}$ by the strip 
$\Gamma=\left\{\tau : -\frac{1}{2}<\tau_1<\frac{1}{2}, \tau_2 >0 \right\}$, while simultaneously setting 
$w=0$ and $w'>0$ in the sum \cite{Polchinski:1985zf}, we can write
$$
    Z = i\,\frac{L_+(2\pi R)V_{23}}{(2\pi)^{26}}\, \int_{\Gamma} \frac{d^2\tau}{\tau_2^{14}}\, |\eta(\tau)|^{-48}\,\sum_{w' = 1}^{\infty}\,
        \int dy\; \exp\left(-\frac{\pi}{\tau_2}\,w'^2R^2(y)\right)\ .
$$
If one sets $R(y)=R$ this expression would be that of the partition function for bosonic strings in Minkowski space 
with one compact direction of radius $R$, which for a critical value of the moduli $R$ has a divergence associated to
a Hagedorn phase transition \cite{Hagedorn:1965st}. Similarly, in the O-plane orbifold we expect to obtain a divergence for a critical value 
of the string center of mass coordinate $y$. To see this expand the Dedekind eta function 
$\eta(\tau)$ in a series and perform the $\tau_1$ integration to obtain
$$ 
Z= i\,\frac{L_+(2\pi R)V_{23}}{(2\pi)^{26}}\,\sum_{w=1,n=0}^{\infty}\,d_n^{\,2}\,\int dy \, 
\int_0^{\infty}\frac{d\tau_2}{\tau_2^{14}}\;\exp\left(-4\pi(n-1)\tau_2-\frac{\pi}{\tau_2}\,w^2R^2(y)\right)\ . 
$$
Let us analyze the $\tau_2$ integral for specific values of $n$. For $n=0$, the integral diverges exponentially as  
$\tau_2\rightarrow \infty$. This is the usual infrared tachyonic divergence of the bosonic closed string. 
Note that the $y$ dependence is irrelevant in what concerns this divergence. For $n\ge 1$ fixed, 
the $\tau_2$ integral diverges for $y\rightarrow 0$. It diverges in the region $\tau_2\rightarrow 0$,
so one might think that this is an ultraviolet divergence. However, this is not the case because the small $\tau_2$
region of the strip $\Gamma$ does not result only from a modular map of the fundamental domain ultraviolet region.
We expect these divergences to be  infrared divergences. An instructive analogy
is to compare  with the case of bosonic string compactified
on a circle in the limit of vanishing radius. In this limit there are winding states that become massless and the divergence 
should be thought in the T-dual picture as a large volume effect.

For positive $y$, the $\tau_2$ integral converges and the partition function can be explicitly written as a Hankel
function of first kind
$$ 
Z=-i\, \frac{L_+(2\pi R)V_{23}}{(2\pi)^{26}} \, 
\sum_{w=1,n=0}^{\infty}\pi\, d_n^{\,2}\,\int dy\,\left(\frac{4(n-1)}{w^2R^2(y)}\right)^{13/2} 
H_{13}^{(1)}\left(4\pi i\sqrt{(n-1)w^2R^2(y)}\right) \ . 
$$
Using the asymptotic expansions for the degeneracy of states and for the Hankel function
$$
d_n\sim n^{-27/4}e^{4\pi\sqrt{n}} \ , \ \ \ \ \ \ 
H_{\nu}^{(1)}(z)\stackrel{|z|\rightarrow \infty}{\sim} \sqrt{\frac{2}{\pi z}} \, e^{\,i(z-\nu\pi/2-\pi/4)} \ , 
$$
we can see that the large $n$ expansion of the integrand in the partition function is dominated by the exponential
term
$$
     e^{\,4\pi \sqrt{n}\left(2-\sqrt{w^2R^2(y)}\right)} \ .
$$
We conclude that, for each winding number $w$, the sum in $n$ diverges when
$$
     y<y_c(w)=\frac{2}{E(w R)^2} \ . 
$$
This behavior of the partition function for large $n$ is of the same type of the Hagedorn divergence of the 
bosonic string compactified on a circle, 
a divergence which can be traced back to the appearance of unstable tachyonic modes in the spectrum 
\cite{Atick:1988si}. 

\section{Hagedorn transition - the chronology warden} \label{warden}

Having found unstable states and related infrared divergences in the string partition function that resemble what happens
in the Hagedorn phase transition, we now suggest how this can protect chronology.

\begin{figure}[t]
\centering\epsfig{file=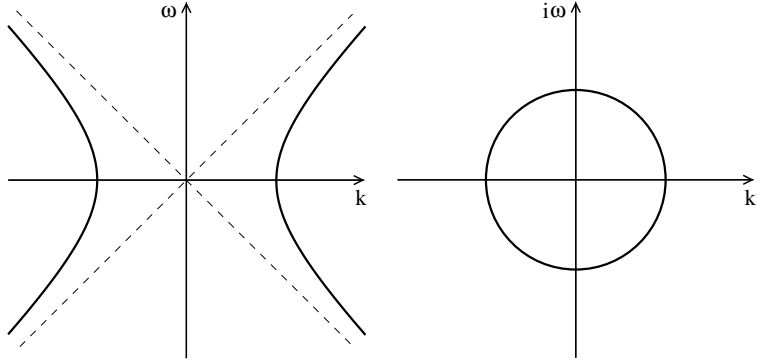,width=10cm}
\caption{\small{Dispersion relation for a tachyon field in Minkowski space.}}
\label{figure10}
\end{figure}

Let us start by recalling the behavior of a tachyon field in Minkowski space. 
The condensation of the field occurs via the growth of low momentum modes, 
as can be seen from the dispersion relation plotted in figure \ref{figure10}. In the case of the bosonic string
one expects that  the closed string tachyon condensation will change the spacetime structure.
For the O-plane orbifold a similar phenomenon occurs. In particular, 
there are winding string states that play an 
important role because the associated field has unstable modes that propagate not only in the bad region, but also a bit inside
the good region. The growth of these modes will then induce a large backreaction in the region where they propagate, triggering a phase
transition. It is therefore natural to conjecture that this phase transition protects chronology.

For simplicity, let us focus on the bosonic string and we shall comment on the extension to the superstring afterwards. 
As usual, we neglect the closed string tachyon. Consider the state with  $m=0$, $w=1$ and $n=\tilde{n}=0$. 
From equation (\ref{stringmass}) we can compute the position where the mass 
$M(y)$ of this field vanishes
\bee
    y = y_c = \frac{2}{ER^2} \ .
\label{ycritical}
\eee
For $y < y_c$ the mass squared of the field becomes negative. Hence we expect a phase transition in this region. This expectation is confirmed by the
existence of on-shell states with imaginary light-cone energy. These states grow significantly  for $y<y_0$, with the quantum number $y_0$ satisfying
$0<y_0 <y_c$, as shown in figure \ref{figure7}.  The singularity developed by the partition function for $y\le y_c$ confirms these results.

The O-plane orbifold phase transition shares some properties with the Hagedorn phase transition. Consider the bosonic string compactified 
on circle of radius $R$. The mass of a string with momentum and winding quantum numbers $(m,w)$ is
\bee
    M^2 = 2\left( n + \tilde{n} - 2\right) + \frac{m^2}{R^2} + (w R)^2 \ .  
\label{standardmass}
\eee
States with quantum numbers $(0,1)$ and no oscillator excitations have $M^2 = -4 + R^2$, becoming massless for  $R=2$.
This is precisely the proper radius of the O-plane orbifold compact direction when the state with the same quantum numbers 
becomes massless. In the usual Hagedorn phase transition the radius $R$ is a moduli. As the radius decreases to $R=2$ the first winding state $w=1$ 
condenses and triggers the phase transition. On the other hand, for the O-plane orbifold the compactification radius varies with the coordinate $y$. 
Since the $w=1$ winding state is the one that penetrates the furthest into the good region, one expects that the condensation of this field will
cause significant changes on the spacetime structure for $y \leq y_c$. In other words, the string fields will develop a large tadpole beyond 
this critical point.

\subsection{Extension to the superstring}

It is straightforward to extend the previous results to the superstring, where one needs to specify the orbifold spin structure. For both
spin structures, we need to recall that,  in the NS-NS sector, the normal ordering for the Virasoro zero-modes
gives the on-shell relation  $\lambda = 2( N + \tilde{N} - 1)$.

Consider first the supersymmetry breaking case
where spacetime fermions have anti-periodic boundary conditions \cite{Rohm:1983aq}. In this case states with even winding number have the usual
supersymmetric GSO projection, while states with odd winding number  have reversed GSO projection. The physical picture is very similar
to the bosonic string just considered, but one does not have the usual bosonic string tachyon. There are winding states that condense 
and whose wave function
also oscillates inside the region $y>0$. These states have $\lambda = -2$, $m=0$ and odd winding number $w$. 
For $w = \pm 1$ the mass $M(y)$ vanishes for $y = y_c = 1/(ER^2)$
and it becomes negative for $y < y_c$. Correspondingly, on-shell states with  $0<y_0< y_c$ have imaginary $p_+$ and grow significantly for $y<y_0$.
The proper radius of the compactification circle at the critical value $y= y_c$ is equal to the superstring Hagedorn critical radius.
Moreover, the partition function has infrared divergences for $y\le y_c$.

The case of supersymmetric boundary conditions is qualitatively different. 
After imposing the standard GSO projection only  states with $\lambda \ge 0$
survive. Now there are no winding states that condense inside the good region.
The contribution from each string field to the partition function will then diverge
for $y_0<0$. However, there will be the usual supersymmetry cancellation between bosons
and fermions.  All unstable states condense in the bad region.
This is consistent with the conjecture put forward in \cite{reviewcosta} to protect
chronology in the presence of closed causal curves, which we now recall. 

Start with the O-plane orbifold in type IIA theory and 
ignore the $y<0$ region. Next uplift the geometry to M-theory and reduce back to type IIA along the orbifold direction $y^-$. 
After performing a T-duality transformation along the eight spectator directions, the result is the geometry
of an O8-plane orientifold \cite{Polchinski:1995df}. 
This is the reason why this orbifold was called the O-plane orbifold. The above duality chain suggests that one should simply
excise the $y<0$ region of the orbifold. Moreover, the quantization of the O8-plane charge was seen to be dual to a quantization 
of the string coupling constant $g_s$ in the orbifold side. Then it was conjectured that such coupling quantization is necessary to restore
unitarity in the O-plane orbifold, lost at arbitrary values of the coupling due to the presence of closed causal curves.

The results of the present paper fit-in nicely with the above conjecture. Our computations are valid at zero string coupling, where we see that the 
string fields condense in the region $y<0$. It is therefore conceivable that the end-point of the supersymmetric O-plane orbifold 
condensation results in the excision of the bad region and in a vacuum expectation value for the dilaton field consistent with 
dual O8-plane charge quantization. 

\section{Concluding remarks}

We believe the results here presented for the O-plane orbifold with supersymmetry breaking boundary conditions 
are general and provide a truly stringy  mechanism to protect chronology. 
Suppose we have an arbitrary spacetime with a Cauchy surface. 
Suppose also that, as the geometry  evolves, we are about to create a closed null curve in the future of such surface. 
Then, when there are closed spacelike curves with proper radius of order of the string length, 
winding string fields become massless and start condensing. There will be a 
Hagedorn phase transition, which protects chronology. In other words, after the phase transition the new vacuum will be
chronological.

Hawking's chronology protection mechanism is based on an ultraviolet effect. He advocates that strong curvature effects
due to a large one loop energy-momentum tensor for the matter fields will create a singularity or will prevent spacetime to curve 
such that non-causal curves do not form. In contrast, the string theory mechanism for chronology protection is an infrared effect,
as it is usually the case for phase transitions. 

Other situations in string theory where the 
appearance of light winding states triggers a phase transition are well known. An interesting follow up of our proposal
is to find the exact end-point of the condensation. Our conjecture is that this
phase transition will remove the bad region from spacetime. Indeed, 
phase transitions associated with light winding string states, that produce such a drastic effect in spacetime
and that lead to a change of topology, are expected to occur in the context of the AdS/Gauge-theory duality 
at finite temperature \cite{Barbon:2004dd}, or in the context of circle compactifications with supersymmetry breaking 
boundary conditions \cite{Adams:2005rb}.
The authors of the latter work argued that, after condensation of the winding string field, the worldsheet theory will 
flow to an infrared trivial fixed point with no target space geometric interpretation. We believe something similar
will happen that excludes the formation of non-causal curves.

In order to check the generality of the mechanism presented herein, interesting laboratories
include other time-dependent orbifolds and other closed string backgrounds, which have already been studied in
relation to closed causal curves in string theory.

\bigskip

\begin{center}
{\bf Acknowledgements}
\end{center}
The authors wish to thank L.~Cornalba for many enlightening discussions.
We would also like to thank F.~Correia and S.~Hirano for useful discussions. 
C.~H., J.~P. and N.~S. are funded by the {\em Funda\c{c}\~ao para a Ci\^encia e Tecnologia} 
fellowships SFRH/BPD/5544/2001, SFRH/BD/9248/2002 and SFRH/BPD/13357/2003, respectively.
This work was also supported by {\em Funda\c c\~ao Calouste Gulbenkian} through {\em Programa de Est\'\i mulo 
\`a Investiga\c c\~ao}, by 
the Marie Curie research grant MERG--CT--2004--511309  and by the FCT 
grants POCTI/FNU/38004/2001 and POCTI/FNU/50161/2003. {\em Centro de F\'\i sica do Porto}
is partially funded by FCT through POCTI programme.

\newpage

\section*{A\ \ \   Wave functions in the O-plane orbifold}

In this appendix we prove several results used in the main text regarding wave functions. 
In the appendix A.1 we construct a complete basis of states in the orbifold geometry and define
a measure for these states. In appendix A.2 we consider on-shell wave functions normalizable on a
surface of constant $y^+$.

\subsection*{A.1\ \ \  Off-shell wave functions and Hilbert space measure}

To define a measure in the Hilbert space we shall regularize the wave functions by working in a box
of volume $V=2\pi R L L_+$. More concretely, we impose the boundary conditions
$$
\begin{array}{c}
\displaystyle{\phi\,(y^-,y,y^+) = \phi\,(y^-+2\pi R,y,y^+)} \spa{0.2}\\
\displaystyle{\phi\,(y^-,y,y^+) = \phi\,(y^-,y,y^+ + L_+)}\spa{0.2}\\
\displaystyle{\phi\left(y^-,-\frac{L}{2},y^+\right) =  \phi\left(y^-,\frac{L}{2},y^+\right)=0 }\ .
\end{array}
$$
Furthermore, it is convenient to work with a basis of functions formed by eigenfunctions of the operator
$$
{\cal H} =  2 {\partial_+}\partial_- + 2Ey \,\left( {\partial_+^{\, 2}} +  \left(w R\right)^2\right) - \partial_y^{\, 2}\ . 
$$
These wave functions can be written in the form
$$
\phi = f(z)\, e^{\,i\left(p_+y^+ + p_-y^-\right)}\ ,
$$
where $p_+ = 2\pi r/L_+$ and $p_-=m/R$, for $r$ and $m$ integers. For fixed winding number $w$, we define $z^3 =K (y_0-y)^3$,  
with the constant $K$ depending on the light-cone energy as 
$$
K =K (p_+)= 2E\,\left(p_+^{\,2} - \left(w R\right)^2\right)\ 
$$ 
and with $y_0$ related to the eigenvalue $-\lambda$ of the operator ${\cal H}$ by 
$$
\lambda = 2p_+p_- + y_0\,K (p_+)\ .
$$
The function $f(z)$ satisfies the Airy differential equation
$$
\left( \partial_z^{\,2} - z \right) f(z) = 0\ ,
$$
subject to Dirichlet boundary conditions at the points 
$$
z_{\pm}^{\,3} = K \left(y_0 \pm\eta\,\frac{L}{2}\right)^3\ ,
$$
where $\eta={\rm sgn}(K)$ and we conveniently chose $z_+>z_-$.
With these boundary conditions the Airy differential equation has solutions only for discrete values of $y_0$,
which we denote as $y_0=y_0(s)$, with $s$ a positive integer.
The spectrum of the operator $\lambda$ is now discrete.

To define the Hilbert space measure in the large volume limit we need to understand the behavior 
of the function $y_0(s)$ for large $L$. Since the function $f(z)$ is a linear combination of the Airy functions
${\rm Ai}(z)$ and ${\rm Bi}(z)$, the boundary condition $f(z_+)=0$ gives
$$
f(z) \,\propto\, {\rm Bi}(z_+)\,{\rm Ai}(z) - {\rm Ai}(z_+)\,{\rm Bi}(z)\ . 
$$
To quantize $y_0$ one imposes the other boundary condition
$$
{\rm Bi}(z_+)\,{\rm Ai}(z_-) = {\rm Ai}(z_+)\,{\rm Bi}(z_-)\ .
$$
The density of states around a fixed value of $y_0$  can be computed
by observing  that in the limit of large $L$  one has $z_+\gg 1$.  Then the above equation 
reduces to ${\rm Ai}(z_-)=0$ because of the exponential growth of the function ${\rm Bi}(z)$ for positive real values of $z$. 
Denoting the zeros of the Airy function Ai by $a_s$, the function
$y_0(s)$ is implicitly defined by
$$
a_s^{\,3} = K \left(y_0(s) -\eta\,\frac{L}{2}\right)^3\ .
$$
Again in the limit of large $L$, one can use the asymptotic expansion for the zeros of the Airy function
$$
a_s = - \left(\frac{3\pi(4s-1)}{8}\right)^{2/3}\Big( 1 + {\cal O}(s^{-2})\Big)
$$
to determine $y_0(s)$. Two consecutive zeros will then give the following difference between discrete values of $y_0$
$$
\delta y_0 \equiv y_0(s+1) - y_0(s) \simeq  - \pi \left[ \,K\, \frac{3\pi(4s-1)}{8}\, \right]^{-1/3}\ ,
$$ 
which can be written as a function of $y_0$
$$
\delta y_0 \simeq -\pi \left[ \,K \left(\eta\,\frac{L}{2} - y_0\right) \, \right]^{-1/2}\ .
$$
Since $\delta y_0 \rightarrow 0$ for $L\rightarrow +\infty$ we can move from sums to integrals 
$$
\sum_s F \left(y_0(s)\right)\ \ \  \rightarrow\ \ \ \int_{-\frac{L}{2}}^{\frac{L}{2}} dy_0\, \rho(y_0)\, F(y_0)\ ,
$$
with measure $\rho = \sum_s \delta\left(y_0 - y_0(s)\right) \simeq 1/|\delta y_0|$ given by
$$
\rho(y_0,p_+) \simeq \frac{1}{\pi}\,\left[ \,K(p_+) \left(\eta\,\frac{L}{2} - y_0\right) \, \right]^{1/2}\ .
$$
Notice that there may be other states with $y_0$ outside the above region of integration, but these are
negligible for  large $L$.

The normalized basis of functions in the large volume limit is
$$
    \phi_{p_+,y_0,m}(\vec{y}) = \frac{|K(p_+)|^{1/3}}{\sqrt{2\pi RL_+ \rho(y_0,p_+)}}\,\Ai(z)\,e^{\,i\left(p_+y^+ + \frac{m}{R}y^-\right)} \ ,
$$
satisfying the orthogonality and completeness conditions
$$
\begin{array}{c}
\displaystyle{\langle {p_+,y_0,m} | {p_+',y_0',m'} \rangle 
        = \frac{2\pi}{L_+}\,\delta(p_+-p_+') \, \frac{1}{\rho(y_0,p_+)}\,\delta(y_0-y_0') \, \delta_{m,m'} \ ,} \spa{0,3}\\
\displaystyle{\sum_m\, \frac{L_+}{2\pi}\int_{-\infty}^{\infty} dp_+ \int_{-\frac{L}{2}}^{\frac{L}{2}} dy_0\,\rho(y_0,p_+)\; 
        \phi_{p_+,m,y_0}^*(\vec{y}\,')\, \phi_{p_+,m,y_0}(\vec{y}) 
        = \delta\left(\vec{y}-\vec{y}\,'\right)} \ .
\end{array}
$$

It is important to have an independent check of the above measure. In the reminder of this appendix we provide that
check. We shall start from the canonical partition function for a particle and derive the result obtained independently in the next
appendix using  the path integral formalism. This involves a Poisson summation of the Ka\l u\.{z}a-Klein quantum number $m$ and a rewriting of the measure 
$\rho(y_0,p_+)$ as an integral over a variable $y$. From the path integral computation one sees that this integration variable should 
be interpreted as the particle average position along the $y$-direction. A similar calculation can be done for the string
partition function.

We start with the canonical particle partition function written as
$$
Z=i\int_0^{\infty} \frac{dl}{2l}\,\sum_m\, \frac{L_+}{2\pi}\int_{-\infty}^{\infty} dp_+ \int_{-\frac{L}{2}}^{\frac{L}{2}} dy_0\,\rho(y_0,ip_+)\; 
\exp\left[-\frac{l}{2}\left( M^2 - 2ip_+\frac{m}{R} + 2Ep_+^{\,2}y_0\right)\right]
$$
and write the measure $\rho(y_0,ip_+)$ in the integral form
$$
\rho(y_0,ip_+) = \frac{|K(ip_+)|^{1/2}}{2\pi}\,\int_{-\frac{L}{2}}^{\frac{L}{2}}dy\,\frac{\Theta\left[\eta(y-y_0)\right]}{\sqrt{\eta(y-y_0)}}\ ,
$$
where $\Theta$ is the usual Heaviside step function.
The integral over $y_0$ reduces to a half Gaussian integral in the $L\rightarrow\infty$ limit
$$
\int_{-\frac{L}{2}}^{\frac{L}{2}} dy_0\, \frac{\Theta\left[\eta(y-y_0)\right]}{\sqrt{\eta(y-y_0)}} \;e^{\,\frac{l}{2}Ky_0} =
\int_{0}^{\frac{L}{2}+\eta y}\frac{du}{\sqrt{u}}\;e^{\,\frac{l}{2}\left(Ky-|K|u\right)} 
\,\,\rightarrow\,\, \left(\frac{2\pi}{l|K|}\right)^{1/2} e^{\,\frac{l}{2}Ky} \ .
$$
Using this result the partition function becomes
$$
Z=i\int_0^{\infty} \frac{dl}{2l}\,\sum_m\, \frac{L_+}{2\pi}\int_{-\infty}^{\infty} dp_+ \int_{-\frac{L}{2}}^{\frac{L}{2}} \frac{dy}{\sqrt{2\pi l}}\; 
\exp\left[-\frac{l}{2}\left( M^2 - 2ip_+\frac{m}{R} + 2Ep_+^{\,2}y\right)\right]\ .
$$
Then, performing the Gaussian integral over the light-cone energy we have
$$
Z=iL_+\int_{-\frac{L}{2}}^{\frac{L}{2}} dy\, \int_0^{\infty} \frac{dl}{2l}\,\frac{1}{2\pi l}\,
\frac{R}{R(y)}\sum_m\,\exp\left[-\frac{l}{2}\left( M^2 +\frac{m^2}{R^2(y)}\right)\right]\ ,
$$
where $R^2(y)=2EyR^2$ as defined in the main text. The Poisson summation formula
$$
\sum_m e^{-\pi A m^2} = \frac{1}{\sqrt{A}}\,\sum_w e^{-\pi w^2 A^{-1}}\ ,
$$
can be used to obtain the expression for the partition function derived from the 
path integral formalism
$$
Z=iL_+(2\pi R)\sum_w\,\int_{-\frac{L}{2}}^{\frac{L}{2}} dy\, \int_0^{\infty} \frac{dl}{2l}\,\frac{1}{(2\pi l)^{3/2}}\;
\exp\left(-\frac{l}{2}\, M^2 - \frac{\big(2\pi w R(y)\big)^2}{2l}\right)\ .
$$

\subsection*{A.2\ \ \  On-shell wave functions}

We shall consider the light-cone evolution of a generic field configuration defined on a surface
of constant $y^+$. Then, given the  field and its first derivative on this surface, 
the evolution is determined by the equations of motion. We shall address this problem
by considering a basis of normalizable functions on the surface
of constant $y^+$ with simple on-shell evolution.

The basis can be constructed using the orbifold modes $\phi_{p_+,y_0,m}(\vec{y})$,
restricted to a constant $y^+$ surface and with the quantum numbers satisfying the on--shell condition
$$
\lambda = 2p_+p_- + 2E\,y_0 \left(p_+^{\,2} - \left(w R\right)^2\right)\ ,
$$
for $\lambda$ fixed. We shall call these functions $\phi_{y_0,m}(y^-,y)$. Discarding an irrelevant
normalization constant, we define these functions by
$$
\phi_{y_0,m}(y,y^-)  =  K^{1/3}\,\Ai\left(K^{1/3}(y_0-y)\right)\, 
e^{\,i \frac{m}{R}y^-}\ .
$$
Their on-shell evolution to the orbifold spacetime is determined by the on-shell 
value of the light-cone energy $p_+$. 
From the asymptotic behavior of the Airy functions, it follows that
$$
K=2E\left(p_+^{\,2} - \left(w R\right)^2\right)
$$
must be real, otherwise the function $\Ai$ would grow exponentially for $y\rightarrow +\infty$
or for $y\rightarrow -\infty$. Notice that, one can not include the other Airy
function $\rm{Bi}$, because it always grows exponentially at least in one
direction. We conclude that the functions $\phi_{y_0,m}(y^-,y)$ are normalizable
provided  both $y_0$ and $p_+$ are real, or $y_0$ is complex and $p_+$ is purely imaginary.

\begin{figure}[t]
\centering\epsfig{file=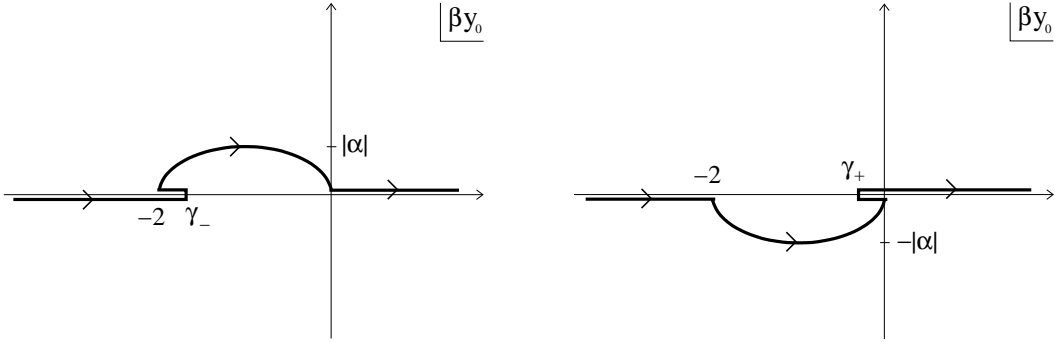,width=14cm}
\caption{\small{Contours in the $\beta y_0$ complex plane for
$|\alpha|<1$. The lines outside the real axis belong to an ellipse with foci at
$\gamma_\pm=-1\pm\sqrt{1-\alpha^2}$. For $|\alpha|=1$ the ellipse becomes a
circle and the contours start, instead, at the center of the circle. For $\alpha=0$ the ellipse degenerates
to the real axis. For $|\alpha|>1$ the foci of the ellipse move to the
complex plane, the left contour starts and finishes at $\beta y_0 \rightarrow +\infty$ and the right
contour at $\beta y_0 \rightarrow -\infty$.}}
\label{figure11}
\end{figure}

The set of solutions to
the on-shell condition with  $K$ real can be parameterized as\footnote{The
  case of $w=0$ can be included by setting $u=w^2 t$ and then
  taking the limit $w\rightarrow 0$ with fixed $t$. In this limit, the ellipse in figure \ref{figure11},
  associated to complex values of $y_0$, degenerates to a parabola.}
$$
\frac{\beta \lambda}{K} = u \ ,\ \ \ \ \ \ \ \ 
\beta y_0=u \pm \alpha\,\sqrt{u^2+2u} \ ,\ \ \ \ \ \ \ \ 
p_+ = \mp \frac{w R}{u}\,\sqrt{u^2 + 2u}\ ,
$$
where $u\in\bb{R}$ and we defined $\alpha= 2mw/\lambda $ and $\beta =E
(2w R)^2/\lambda$. In this parameterization the unstable modes appear for
$-2 < u < 0$. For each value of $u$ there are two 
functions $\phi^{\,\pm}_{u,m}(y^-,y)$, corresponding to the choice of signs in the above parameterization.
In the general case $m\neq 0$, these functions are different because they have different
values of $y_0$. In figure \ref{figure11},  both contours in
the $y_0$ complex plane for on-shell states are shown. In the simpler case
$m=0$, both contours coalesce to the real line and, for each value of $u$, 
the two functions $\phi^{\,\pm}_{u,m}(y^-,y)$ are equal with symmetric values of the light-cone
energy.

Using the above parameterization, the initial value problem reduces to finding the functions
$\tilde{F}^{\pm}(u,m)$    such that
$$
\begin{array}{c}
\displaystyle{F(y,y^-)}= 
\displaystyle{\sum_m \int du \left[ \tilde{F}^+(u,m)\,\phi^{\,+}_{u,m}(y,y^-)
+\tilde{F}^-(u,m)\,\phi^{\,-}_{u,m}(y,y^-)\right]}\ ,\spa{0.6}\\
\displaystyle{F'(y,y^-)} = 
\displaystyle{-iw R \sum_m \int du\,u^{-1}\sqrt{u^2+2u}
\left[\tilde{F}^+(u,m)\,\phi^{\,+}_{u,m}(y,y^-)-\tilde{F}^-(u,m)\,\phi^{\,-}_{u,m}(y,y^-)\right]}\ ,
\end{array}
$$
for any given function $F(y,y^-)$ and its first derivative $F'(y,y^-)$. 
We start by decomposing
the function and its derivative in Fourier modes
$$
\begin{array}{c}
\displaystyle{F(y,y^-)}= 
\displaystyle{\sum_m \int dq \;e^{iqy + i \frac{m}{R}y^-} G_m(q)}\ ,\spa{0.6}\\
\displaystyle{F'(y,y^-)} = 
\displaystyle{\sum_m \int dq \;e^{iqy + i \frac{m}{R}y^-} G'_m(q)}\ .\ \ \ 
\end{array}
$$
Then, by linearity, we can break the problem in two cases:  $F$
generic and $F'=0$; $F'$ generic and $F=0$. Moreover,
we only need to consider each Fourier mode independently.

Firstly, we consider an initial configuration with zero derivative
$$
F(y,y^-)=e^{iqy + i \frac{m}{R}y^-}\ ,\ \ \ \ \ \ \ \ \ \  \ \ \ \ 
F'(y,y^-)=0 \ .
$$
Using the integral representation of the Airy function\footnote{This integral representation is valid 
for complex values of $y_0(u)$,  provided one deforms the real axis integration
contour as 
$s\rightarrow s+i\epsilon\, {\rm sgn}(u)$.}
$$
\phi_{u,m}^{\,\pm}(y,y^-)\, =\, \beta\;e^{\,i \frac{m}{R}y^-} 
\int^{+\infty}_{-\infty} \,\frac{ds}{2\pi} \; 
\exp \left[i s\left(\beta y - u \mp \alpha\sqrt{u^2+2u}\right) - is^3\,\frac{\beta^2u}{3\lambda}\right]\ ,
$$ 
we find that the
corresponding coefficients $\tilde{F}^{\,\pm}_{q}(u,m)$ in the expansion have to satisfy
$$
\begin{array}{l}
\displaystyle{\delta(q-\beta s)} \,=\,
\displaystyle{\int\,\frac{du}{2\pi} \; e^{-iu\left(s+\frac{s^3\beta^2}{3\lambda}\right)} 
\left[e^{-is\alpha\sqrt{u^2+2u}}\,\tilde{F}^{\,+}_{q}(u,m)
+e^{\,is\alpha\sqrt{u^2+2u}}\,\tilde{F}^{\,-}_{q}(u,m)\right]\ ,}
\spa{0.7}\\
\displaystyle{0} =
\displaystyle{\int\,du\, \,u^{-1}\sqrt{u^2+2u}
\; e^{-iu\left(s+\frac{s^3\beta^2}{3\lambda}\right)} \left[e^{-is\alpha\sqrt{u^2+2u}}\,\tilde{F}^{\,+}_{q}(u,m)
-e^{\,is\alpha\sqrt{u^2+2u}}\,\tilde{F}^{\,-}_{q}(u,m)\right]\ .}
\end{array}
$$
Secondly, we choose the initial condition
$$
F(y,y^-)=0\ ,\ \ \ \ \ \ \ \ \ \  \ \ \ \ 
F'(y,y^-)=e^{iqy + i \frac{m}{R}y^-} \ ,
$$
and find the following equations for the corresponding $\tilde{F}'^{\,\pm}_{\;q}(u,m)$
$$
\begin{array}{l}
\displaystyle{0} =
\displaystyle{\int\,du\; e^{-iu\left(s+\frac{s^3\beta^2}{3\lambda}\right)} 
\left[e^{-is\alpha\sqrt{u^2+2u}}\,\tilde{F}'^{\,+}_{\;q}(u,m)
+e^{\,is\alpha\sqrt{u^2+2u}}\,\tilde{F}'^{\,-}_{\;q}(u,m)\right]\ ,}
\spa{0.7}\\
\displaystyle{\delta(q-\beta s)}\, =\,
\displaystyle{-iw R \int\,\frac{du}{2\pi} \,u^{-1}\sqrt{u^2+2u}
\; e^{-iu\left(s+\frac{s^3\beta^2}{3\lambda}\right)}} \times
\spa{0.6}\\
\ \ \ \ \ \ \ \ \ \ \ \ \ \ \ \ \ \ \ \ \ \ \ \ \ \ 
\displaystyle{\times\left[e^{-is\alpha\sqrt{u^2+2u}}\,\tilde{F}'^{\,+}_{\;q}(u,m)
-e^{\,is\alpha\sqrt{u^2+2u}}\,\tilde{F}'^{\,-}_{\;q}(u,m)\right]\ .}
\end{array}
$$

Finally, the expansion for the generic function $F$ with derivative
$F'$ has coefficients
$$
\tilde{F}^\pm(u,m) = \int dq
\left[ G_m(q)\,\tilde{F}^{\,\pm}_q(u,m) + G'_m(q)\,\tilde{F}'^{\,\pm}_{\;q}(u,m)\right]\ .
$$
Unfortunately, we do not know if the basis of functions is complete because we 
were not able to compute the functions $\tilde{F}^{\,\pm}_q(u,m)$
and $\tilde{F}'^{\,\pm}_{\;q}(u,m)$ in the general case.
However, in the case of functions independent of $y^-$, so that
we only need to consider wave functions with $m=0$, we showed that the basis
is complete for $\lambda>0$ and that it spans very generic functions for
$\lambda<0$, as explained below. The latter case is the most 
relevant case for the results of this paper, since it includes the winding 
states that were argued to protect chronology in the O-plane orbifold.  

For $m=0$ the on-shell relation becomes $\lambda = y_0 K$, and therefore
reality of $K$ implies that $y_0$ is also real. In the $u$ parameterization 
this corresponds to setting $\alpha=0$ and the above formulas simplify
considerably. In particular, defining
$$
\begin{array}{c}
\displaystyle{\tilde{F}_q^{\,+}(u,0)= \tilde{F}_q^{\,-}(u,0) = 
\frac{1}{2}\, \tilde{F}_q(u)\ ,}
 \spa{0.5}\\
\displaystyle{\tilde{F}'^{\,+}_{\;q}(u,0) = - \tilde{F}'^{\,-}_{\;q}(u,0) =
\frac{iu}{2w R\sqrt{u^2 + 2u}}\,\tilde{F}_q(u)\ ,}
\end{array}
$$
one is left to find $\tilde{F}_q$ such that
$$
\delta (q-\beta s) \,=\, \int\,\frac{du}{2\pi} \;
e^{-iu\left(s+\frac{s^3\beta^2}{3\lambda}\right)}\, 
\tilde{F}_q(u)\ .
$$
For $\lambda > 0$ this equation has solution
$$
\tilde{F}_q(u) = \frac{1}{\beta} \left( 1 + \frac{q^2}{\lambda}\right) \,
e^{\,i\,\frac{u}{\beta}\,\left( q +  \frac{q^3}{3\lambda}\right)}\ ,
$$
showing the completeness of the basis for  $\lambda > 0$ (and $\lambda=0$
by adding a mass regulator). When $\lambda<0$, the function $s+\beta^2s^3/3\lambda$
is not injective in the interval $|\beta s|< 2\sqrt{-\lambda}\,$. Therefore,
the above explicit solution for $\tilde{F}_q$ is only valid for
$|q|>2\sqrt{-\lambda}\,$. For $|q|<2\sqrt{-\lambda}\,$, the integral equation
for $\tilde{F}_q$ does not have a solution, showing that the basis is not
complete. The subspace of functions generated by this basis is the set of
functions whose Fourier transform $G_0(q)$ is a function of the particular
combination $q+q^3/3\lambda$. Hence, the Fourier coefficients with 
$|q|>\sqrt{-3\lambda}$ can be fixed arbitrarily, but also determine the
remaining low momenta Fourier modes. We conclude that we can construct well localized
perturbations, with some constraint on their long range decay properties
(this can be deduced from the constraints on the pole structure of
the Fourier transform). 
An interesting question is if one can complete the basis or, 
on the other hand, if the above restrictions are necessary for
consistent  evolution in the presence of closed causal curves.

\section*{B\ \ \  Path integral formalism}

In this appendix we compute the particle and string partition functions using the path integral formalism. The expressions
we obtain can also be derived from the canonical formalism using the method described in the previous appendix. They serve
as a check of the measure in the Hilbert space and provide a physical interpretation for the integration variable $y$.

Before we start the computations it is convenient to write the orbifold
identifications more compactly as 
$$
    \Omega{(R,\Delta)}\, \vec{x}\equiv  e^{-2\pi\Delta {\mathcal J}}\vec{x}+2\pi R\, \vec{T}{(\pi\Delta)} \ ,
$$
where we defined the column vector $\vec{T}(a)$ and the nilpotent matrix ${\cal J}$ as
$$  \vec{T}(a) \equiv\left( \barr{c} 1 \\
\displaystyle{-a}
\\ \displaystyle{\frac{2}{3}a^2} \earr \right)  \ ,
\ \ \ \ \ \ \ \ \ 
{\cal J}\equiv\left(\barr{c} 0 \ 0 \ 0 \\
1 \ 0 \ 0 \\ 0 \ 1 \ 0 \earr \right) \ .  
$$
A straightforward computation shows that the action of the orbifold group on
a vector $\vec{x}$ satisfies
$$
\vec{x}_n \equiv \Omega^{n}{(R,\Delta)}\,\vec{x}=\Omega{(nR,n\Delta)}\,\vec{x}\ .
$$

\subsection*{B.1\ \ \  Particle partition function}

The one-loop vacuum energy for a field of mass $M$ is given by the particle partition function
$$
Z=\int_0^{\infty} \frac{dl}{2l}\,\int {\cal D} \vec{X} \;
\exp\left[ - \frac{l}{2}\,M^2 - \frac{1}{2l}\,\int_0^1 \dot{X}^2d\tau\right]
$$
where the functional integral is done over periodic trajectories. Working in covering space, 
these trajectories are periodic up to the orbifold identification
$$
\vec{X} (\tau + l) = e^{\,-2\pi w \Delta {\mathcal J}} \vec{X}(\tau) + 2 \pi w R\, \vec{T}(\pi w\Delta) \ ,
$$
where $w$ is the trajectory winding number. 
A general path can then be expanded in a Fourier series as
$$
\vec{X}(\tau) = e^{\,-2\pi w \Delta\tau{\mathcal J}} \,\sum_n \vec{X}_n \, e^{\,2\pi i n \tau}
        + 2\pi w R\tau\,\vec{T} \left(\pi w \Delta\tau\right) \ ,
$$
where $\vec{X}_n=\vec{X}_{-n}^{\,*}$ so that  $\vec{X}$ is real. Replacing this expansion in the action the contributions
from zero-modes and from quantum fluctuations decouple, so the partition function has the form
$$
Z=\sum_w\,\int_0^{\infty} \frac{dl}{2l}\,{\cal Z}_{\rm qu}\,{\cal Z}_{\rm cl}\;\exp\left( -\frac{l}{2}\,M^2\right)\ . 
$$

The contribution from quantum fluctuations is given by\footnote{The factor of $2^3$ in the measure appears because the coefficients associated to
real normalized modes are $\sqrt{2}\,{\rm Re} (\vec{X}_n)$ and $\sqrt{2}\,{\rm Im} (\vec{X}_n)$.}
$$
{\cal Z}_{\rm qu} = \int \left(\prod_{n=1}^{\infty} 2^3d\vec{X}_nd\vec{X}_n^{\,*}\right)
\;\exp\left( -\sum_{n=1}^{\infty}\vec{X}_n^{\,\dagger} M_n \vec{X}_n\right)
= \prod_{n=1}^{\infty} \frac{(2\pi)^3}{{\rm det} M_n}\ ,
$$
where the matrix $(M_n)_{\mu\nu}$ is
$$
M_n = \frac{(2\pi n)^2}{l}\,\left(
\begin{array}{ccc}
\displaystyle{\left(\frac{w \Delta}{n}\right)^2} 
& \displaystyle{-\frac{2w\Delta}{n}\,i} 
& \displaystyle{-1}\spa{0.4}\\
\displaystyle{\frac{2w\Delta}{n}\,i} 
& \displaystyle{1} & 0 \spa{0.4}\\
\displaystyle{-1} & 0 & 0 
\end{array}
\right)\ .
$$
The determinant is therefore equal to that in the Minkowski space computation and one obtains after zeta function
regularization ${\cal Z}_{\rm qu} = i(2\pi l)^{-3/2}$.

Next we consider the contribution from the zero-modes 
$$
{\cal Z}_{\rm cl} = \int_{\cal F} d\vec{X}_0 \;
\exp\left[ - \frac{(2\pi w R)^2 E}{l}\,\left( X_0 + \frac{E}{2}\,\left(X_0^-\right)^2\right)\right]\ ,
$$
where the integration is over the orbifold fundamental domain. Moving to the $\vec{y}$-coordinates this reduces
to
$$
{\cal Z}_{\rm cl} = \int dy^-dy dy^+\;  
\exp\left(- \frac{\big(2\pi w R(y)\big)^2}{2l}\right)\ .
$$
Integrating over the $y^+$ and $y^-$ directions, we obtain the final answer for the
particle partition function
$$
Z=iL_+(2\pi R)\sum_w\,\int_{-\infty}^{\infty}dy\, \int_0^{\infty} \frac{dl}{2l}\,\frac{1}{(2\pi l)^{3/2}}\;
\exp\left(-\frac{l}{2}\, M^2 - \frac{\big(2\pi w R(y)\big)^2}{2l}\right)\ .
$$

\subsection*{B.2\ \ \  String partition function}

We now compute the bosonic string partition function
$$
    Z = \int {\cal D} \vec{X} \,  {\cal D} {\cal \gamma} \, e^{-S\left[\vec{X},\gamma\right]} \ ,
$$
where the Euclidean worldsheet action is
$$
    S = \frac{1}{4\pi} \int d^2 \sigma \sqrt{\gamma}\,\gamma^{ab} \partial_a \vec{X} \cdot \partial_b \vec{X} \ .
$$
The dot denotes the inner product $\vec{A}\cdot\vec{B} = A^T G B$, with $G$ the target space metric. Using the symmetries
of the action we fix the metric on the torus $\gamma_{ab}$ to be given by $ds^2 = |d\sigma_1 + \tau d\sigma_2|^2$, 
with modular parameter $\tau = \tau_1 + i\tau_2$. This gauge fixing introduces ghost contributions which will be added
in the end. With this choice of metric, and following the notation introduced above, 
the orbifold identifications defining the $(w_1,w_2)$-twisted sector are
\bena
    &&\vec{X} (\sigma_1 + \pi,\sigma_2) = e^{-2\pi w_1\Delta  {\cal J}} \vec{X} (\sigma_1,\sigma_2) 
                                        + 2 \pi Rw_1 \vec{T} (\pi w_1\Delta ) \ , \\
    &&\vec{X} (\sigma_1,\sigma_2 + \pi) = e^{-2\pi w_2\Delta  {\cal J}} \vec{X} (\sigma_1,\sigma_2) 
                                        + 2 \pi Rw_2 \vec{T} (\pi w_2\Delta ) \ .
\eena
Functions that obey these conditions can be written as a Fourier series
$$
    \vec{X} = e^{-2\Delta v {\cal J}} \sum_{n_1,n_2} \vec{X}_{n_1,n_2} \, e^{{2 i(n_1\sigma_1 + n_2\sigma_2)}} 
            + 2R v \vec{T} (\Delta v) \ ,
$$
where $v=w_1\sigma_1+w_2\sigma_2$ and reality of $\vec{X}$ requires $\vec{X}_{-n_1,-n_2}=\vec{X}_{n_1,n_2}^{\,*}$. 
The $\vec{X}_{0,0}$ are zero-modes while the other $\vec{X}_{n_1,n_2}$ describe quantum fluctuations. 

The gauge fixed partition function without the ghost contribution has the explicit form
$$
   Z = \int_{\cal F} d^2\tau
       \int {\cal D} \vec{X} \, \exp \left\{ \frac{1}{4\pi} \int d^2 \sigma \,\frac{1}{\tau_2} 
       \left( |\tau|^2 \partial_1 \vec{X} \cdot \partial_1 \vec{X} 
               - 2 \tau_1 \partial_1 \vec{X} \cdot \partial_2 \vec{X} 
               + \partial_2 \vec{X} \cdot \partial_2 \vec{X} \right) \right\} \ ,
$$
where the $\tau$ integration is over the fundamental domain of the torus ${\cal F}$.
To evaluate the action of the above Fourier expansion of the fields it is convenient to consider the zero-modes and quantum fluctuations separately. 
We include the term $2R v \vec{T}(\Delta v)$ in the zero-mode contribution 
to the partition function. Start with the quantum fluctuations, for which we follow the analysis of \cite{Seiberg:2002,Fabinger:2002kr}. 
In this case the action has contributions of the type
$$
\begin{array}{rcl}
   \displaystyle{\partial_a \vec{X} \cdot \partial_b \vec{X}} &=& 
    \displaystyle{  \sum_{n_1,n_2\neq 0,0} \left(  2i n_a \, e^{-2\Delta v {\cal J}} - 2w_a\Delta  {\cal J} \, e^{-2\Delta v {\cal J}} \right) 
      \vec{X}_{n_1,n_2}\, e^{2 i (n_1\sigma_1+n_2\sigma_2)}}\spa{0.5}
\\ && \displaystyle{\cdot
      \sum_{n_1,n_2\neq 0,0} \left(  2i n_b \, e^{-2\Delta v {\cal J}} - 2w_b\Delta  {\cal J} \, e^{-2\Delta v {\cal J}} \right) 
      \vec{X}_{n_1,n_2}\, e^{2 i (n_1\sigma_1+n_2\sigma_2)}}  \ .
\end{array}
$$
The structure of the various terms is such that 
after integrating over $\sigma_1$ and $\sigma_2$ we are left only with terms which, for each mode $(n_1,n_2)$, can be written as 
$$
\vec{X}^{\,\dagger}_{n_1,n_2}\left( \sum_{r,s=0}^2 C_{rs}({\cal J}^r)^T G\, {\cal J}^s \right) \vec{X}_{n_1,n_2} \ .
$$
The path integral requires us to do the integration over the Fourier modes. These are Gaussian integrals and for each mode
can be obtained from the determinant of a matrix $(M_{n_1,n_2})_{\mu\nu}$ of the form
$$
  M_{n_1,n_2} = \sum_{r,s=0}^2 C_{rs}({\cal J}^r)^T G \,{\cal J}^s =
   \left( \begin{array}{ccc}
      M_{--} & M_{-x} & M_{-+} \\
      M_{x-} & M_{xx} & 0 \\
      M_{+-} &   0    & 0 \\
   \end{array} \right) \ .
$$
It is then easy to see that  the terms that give a contribution to the determinant of $M_{n_1,n_2}$ have $r=s=0$. These
terms only appear for $w_1=w_2=0$, so quantum fluctuations are insensitive to the winding of the string. 
The same happened in the computation of the partition functions for the null-boost and null-brane orbifolds presented in 
\cite{Seiberg:2002,Fabinger:2002kr}. Including the contribution of ghosts, the result is then 
the same as in Minkowski space, namely
$$
   {\cal Z}_{\rm qu}(\tau) = \frac{i}{\tau_2}\,(2\pi\sqrt{\tau_2})^{-3} |\eta(\tau)|^{-2} \ .
$$
This is just the partition function of three free bosons plus ghosts.

Next we evaluate the zero-mode contribution to the partition function. When $n_1=n_2=0$, $\vec{X}$ simplifies to
$$
\vec{X} = e^{-2\Delta v {\cal J}} \vec{x} + 2R v \vec{T} (\Delta v)\ ,
$$
where we defined $\vec{X}_{0,0}\equiv \vec{x}$. The string center of mass $\vec{X}$ does not coincide with the zero-mode $\vec{x}$. 
However, changing to the $\vec{y}$-coordinates, one can check that the $y$ component of the zero-mode coincides with 
the string center of mass component $Y$. This is an important point since, in the main text, we interpreted the integration 
variable $y$ as the string center of mass. A straightforward computation shows that
$$
   \partial_a \vec{X} \cdot \partial_b \vec{X} = 8ER^2 w_a w_b \left( x + \frac{E}{2} (x^-)^2 \right) 
                                               = 8ER^2 w_a w_b \,y \ ,
$$
which  yields an action
$$
   S = \frac{2\pi ER^2}{\tau_2} \, y \, \bb{T}\bar{\bb{T}} \ , \ \ \ \ \ \ \ \ \ \ \ \ \bb{T} = w_1 \tau - w_2 \ .
$$
Adding the oscillators contribution from the transverse directions, the full partition function is then
$$
   Z = i\,\frac{L_+(2\pi R)V_{23}}{(2\pi)^{26}}\, \int_{\cal F} \frac{d^2\tau}{\tau_2^{14}}\,|\eta(\tau)|^{-48}\,\sum_{w_1,w_2}\, 
       \int_{-\infty}^{\infty} dy\; \exp\left(-\frac{\pi}{\tau_2}\,R^2(y)\,\bb{T}\bar{\bb{T}} \right)\ . 
$$

\end{document}